\documentclass[twocolumn,aps,prl,superscriptaddress,nofootinbib]{revtex4-2}

\usepackage[dvipsnames,svgnames]{xcolor}
\usepackage{slashed}
\usepackage{tikz-cd}
\usepackage{tikz}
\usepackage{relsize}
\usepackage{amsmath}
\usepackage{amsfonts}
\usepackage{bbold}
\usepackage{amssymb}
\usepackage{graphicx}
\usepackage{mathtools}
\usepackage{stmaryrd}
\usepackage{bbm}	
\usepackage{mathtools,leftindex,tensor,mhchem}
\usetikzlibrary{decorations.pathreplacing,calligraphy,decorations.markings}
\usepackage[colorlinks=true,allcolors=blue]{hyperref}
\hypersetup{breaklinks=true}
\usepackage{amsthm}
\usepackage{cleveref} 
\newtheoremstyle{mystyle}
    {\dimexpr\topsep/2\relax} 
    {\dimexpr\topsep/2\relax} 
    {}          
    {\parindent}          
    {\bfseries} 
    {.}         
    {.5em}      
    {}          
\theoremstyle{mystyle}
\newtheorem{lemma}{Lemma}
\newtheorem{defn}{Definition}

\newtheorem{thm}{Theorem}

\definecolor{whampdocolor}{HTML}{FCDE70}
\iftutex
  \usepackage{unicode-math}
  \setmathfontface\altgrfont{AntykwaTorunskaMed-Italic.otf}[Scale=MatchLowercase]
  \newcommand{\kronecker}{\altgrfont{δ}}
\else
  \usepackage[T1]{fontenc}
  \usepackage[utf8]{inputenc} 
  \usepackage{amsmath}
  \DeclareSymbolFont{altgr}{OML}{antt}{m}{it}
  \DeclareMathSymbol{\kronecker}{\mathord}{altgr}{"0E}
\fi

\tikzset{baseline={([yshift=-.5ex]current bounding box.center)}}

\newcommand{\splitspace}[5]{
	\begin{scope}[shift={(#1)}]
		\draw (0,0) -- (1,1);
		\draw (0,0) -- (-1,1);
		\draw (0,0) -- (0,-1);
        \draw (-1.4,1) node {\small #2};      
        \draw (1.4,1) node {\small #3};      
        \draw (0.4,-1) node {\small #4};
		\draw (0,0.8) node {\small #5};
	\end{scope}
}

\newcommand{\fusionspace}[5]{
	\begin{scope}[shift={(#1)}]
		\draw (0,0) -- (1,-1);
		\draw (0,0) -- (-1,-1);
		\draw (0,0) -- (0,1);
        \draw (-1.4,-1) node {\small #2};      
        \draw (1.4,-1) node {\small #3};      
        \draw (0.4,1) node {\small #4};
		\draw (0,-0.8) node {\small #5};
	\end{scope}
}

\newcommand{\QCA}[1]{
	\begin{scope}[shift={(#1)}]
		\fill[gray] (0.1,0.1) rectangle (0.9,0.9);
		\draw[thick] (0.4,-0.5)--(0.4,0);
		\draw[thick] (0.4,1)--(0.4,1.5);
		\draw[thick, blue] (0.6,0)--(0.8,-0.4);
		\draw[thick, blue] (0.6,1)--(0.8,1.4);
	\end{scope}
}

\begin{document}
\title{Anomalous matrix product operator symmetries and 1D mixed-state phases}
\author{Xiao-Qi Sun}\email{xiaoqi.sun@mpq.mpg.de}
\affiliation{Max Planck Institute of Quantum Optics, Hans-Kopfermann-Stra{\ss}e 1, D-85748 Garching, Germany}
\begin{abstract}

Generalized symmetries have emerged as a powerful organizing principle for exotic quantum phases. However, their role in open quantum systems, especially for non-invertible cases, remains largely unexplored. We address this by applying a unified tensor-network framework for mixed states with fusion categorical symmetry, which encompasses both invertible and non-invertible ones represented as matrix product operators,  and reveals novel quantum phases unique to the open-system setting through the lens of quantum anomalies. In contrast to pure states, where anomalies forbid symmetric short-range correlated phases in one dimension, we construct a broad class of renormalization fixed-point mixed states with zero correlation length given arbitrary strong anomalous fusion categorical symmetry. These states, representing nontrivial mixed-state phases of matter, cannot be efficient prepared via local quantum channels, indicating anomaly-enforced long-range entanglement in the absence of local correlations. Despite this obstruction, we further provide constructions of measurement-enhanced quantum circuits to prepare all these constructed states, offering a practical way to realize and probe anomalous generalized symmetries in open quantum systems.

\end{abstract}
\maketitle
\emph{Introduction.--} While the conventional notion of global symmetry includes only space-time and internal symmetries,  generalizations of symmetries have played an increasingly important role in modern condensed matter and high-energy physics. One notable generalization involves non-invertible symmetries (see \cite{Mcgreevy2023,Cordova2022, Brennan2023, Shao2023,Schafer2024} for reviews), which go beyond the unitary and anti-unitary symmetries derived from Wigner's theorem. In recent years, it has become evident that such symmetries arise naturally in a variety of quantum many-body systems.  A notable class of non-invertible symmetries arises from self-duality, where the physical system remains invariant under a non-invertible transformation at the critical point of a phase transition, as exemplified by the Kramers–Wannier self-duality at the (1+1)D Ising critical point. While generalized symmetries provide a powerful organizing principle~\cite{Mcgreevy2023} for both gapped and gapless phases of matter in closed systems, their role in open quantum systems remains largely unexplored especially for the non-invertible cases. One fundamental problem is to examine constraints from generalized symmetries on long-distance universal physics of many-body mixed states and uncover nontrivial mixed states beyond the pure-state understanding. 

In this Letter, we tackle this problem from the perspective of quantum anomaly. The quantum anomaly of symmetries in open quantum systems has richer structures as the interaction with environment distinguishes two types~\cite{Buvca2012,Albert2014,deGroot2022,Ma2023a}: a strong one where the system's symmetry charge is conserved exactly, and a weak one, where the symmetry charge can be exchanged with the environment but remains conserved on average. For conventional symmetries, it was shown~\cite{Lessa2025a} that the 't Hooft anomaly~\cite{Hooft1980} of strong symmetries can enforce long-range entanglement~\cite{Chen2010} for 1D mixed states. Related mechanisms have also been explored in higher dimensions~\cite{Lessa2025a,Lessa2025,Hsin2025}.  Here we consider the quantum anomalies of generalized symmetries described by unitary fusion categories, encompassing both invertible and non-invertible cases, for 1D mixed states within a unified framework of matrix product operators (MPOs)~\cite{Bultinck2017,Csahinouglu2021,Lootens2021}. We establish a fundamental obstruction to efficient preparation via local quantum channels for mixed states with strong anomalous generalized symmetry. To elucidate the implications in the classification of mixed state phases of matter, we explicitly construct a broad class of zero-correlation-length mixed states at renormalization fixed points with such strong anomalous generalized symmetries, indicating nontrivial short-range correlated phases with anomaly-enforced long-range entanglement. Finally, we show that all these constructed states can be prepared efficiently by quantum circuits enhanced by measurement and feedforward, opening up an avenue to realize and probe anomalous generalized symmetries in open quantum systems.

\emph{Anomalies and matrix product operator symmetries.--} More general than a group, we consider symmetries forming algebras, represented by MPOs. We define these as MPO symmetries, which encompass non-invertible symmetries and/or anomalous symmetries: 
\begin{defn} 
An MPO symmetry is a representation of a finite-dimensional complex algebra $\mathcal{R}$ (with basis elements $\{\mathcal{L}_a\}_{a\in \mathcal{B}}$ and structure constants $N_{ab}^c$) on Hilbert space $\mathcal{H}^{\otimes L}$ for all $L$, where $\mathcal{H}$ is finite-dimensional, realized by translationally invariant (TI) normal MPOs:
\begin{equation}
O_a^{(L)}=\sum_{\{i\},\{j\}}\text{tr}[T_a^{i_1j_1}T_a^{i_2j_2}...T_a^{i_L j_L}]|i_1 i_2...i_L\rangle\langle j_1 j_2...j_L|,
\end{equation}
with $T_a^{ij}$ being finite-dimensional matrices and these MPOs satisfy
\begin{equation}
 O_a^{(L)}O_b^{(L)}=\sum_c N_{ab}^c O_c^{(L)},
\end{equation}
independent of $L$.
\end{defn}

\noindent We refer to $O_a^{(L)}$ as a normal MPO when $T_a$ is a normal tensor~\cite{Cirac2017}, i.e., (1) the matrices $T_{a}^{ij}$ have no nontrivial common invariant subspace, and (2) the transfer matrix $\sum_{ij}(T_{a}^{ij})^*\otimes T_{a}^{ij}$ has a unique eigenvalue of largest magnitude. The requirement of $L$-independence constrains $N_{ab}^c$ to be non-negative integers~\cite{Bultinck2017}. 

Given an MPO symmetry, it is useful to study MPO symmetric states and, in the spirit of the Lieb-Schulz-Mattis theorem~\cite{Lieb1961,Oshikawa2000, Hastings2004}, define the anomalous MPO symmetry from the absence of symmetric and short-range correlated ones. We restrict on TI matrix product states (MPSs)~\cite{Fannes1992}:
\begin{equation}
|\psi^{(L)}(A)\rangle=\sum_{\{i\}}\text{tr}[A^{i_1}A^{i_2}...A^{i_L}]|i_1,i_2,...,i_L\rangle,
\end{equation}
with $A$ a rank-3 tensor, $A^i$ being finite dimensional matrices, and the symmetry condition is as follows:
\begin{defn} 
A TI MPS $|\psi^{(L)}(A)\rangle$ has the MPO symmetry generated by $\{O_a^{(L)}\}_{a\in\mathcal{B}}$ if for all $a$ and $L$
\begin{equation}
O_a^{(L)}|\psi^{(L)}(A)\rangle=\lambda_a^{(L)}|\psi^{(L)}(A)\rangle,
\label{eq:symm}
\end{equation}
and $\mathcal{L}_a\rightarrow \lambda_a^{(L)}|_{a\in\mathcal{B}}$ is a 1D representation of $\mathcal{R}$ over $\mathbb{C}$.
\end{defn}
\noindent Now we define the anomalous MPO symmetry. Consider a family of MPO symmetries $\{O_a^{(L)}\otimes \mathbb{1}_r^{\otimes L}\}_{a\in \mathcal{B}}$ where each site is augmented with an ancillary space of dimension $r$ that transforms trivially under the symmetry. The notion of anomalous symmetry should be robust upon adding these ancillas:
\begin{defn} 
An MPO symmetry $\{O_a^{(L)}\}_{a\in\mathcal{B}}$ is anomalous if there is no TI normal MPS that is symmetric under the MPO symmetry $\{O_a^{(L)}\otimes \mathbb{1}_{r}^{\otimes L}\}_{a\in\mathcal{B}}$ for all  finite $r$.
\end{defn}
\noindent Normal MPSs are generated by normal tensor $A$ satisfying similar standard conditions as we have discussed for $T_a$. Physically, they represent short-range correlated states. For unitary symmetries, the quantum anomaly remains stable even when ancillas admit nontrivial onsite symmetry transformations. While our definition restricts ancillas to transform trivially, it suffices for later discussions on the preparation complexity of mixed states. We now prove a useful property of non-anomalous MPO symmetries:

\begin{lemma}
    For any non-anomalous MPO symmetry $\{O_a\}_{a\in \mathcal{B}}$ with a symmetric normal MPS $O_a^{(L)}|\psi^{(L)}(A)\rangle=\lambda_a^{(L)}|\psi^{(L)}(A)\rangle$, $\lambda_a^{(L)}$ must be periodic in $L$ and there exists an $L$ such that $\lambda_a^{(L)}=n_a$ for all $a$, where $n_a$ is a non-negative integer. 
    \label{lemma:integral}
\end{lemma}

\emph{Proof:}  For any given $a$, if $\lambda_a^{(L)}=0$ for all $L$, then $n_a=0$. Otherwise $O_a^{(L)}|\psi^{(L)}(A)\rangle$ and $|\psi^{(L)}(A)\rangle$ are proportional TI MPSs for all $L$. Applying the fundamental theorem~\cite{Cirac2017}, we must have only one normal tensor, which can be chosen to be $A$, as basis of normal tensors for $O_a^{(L)}|\psi^{(L)}(A)\rangle$, such that  $O_a^{(L)}|\psi^{(L)}(A)\rangle=\left(\sum_{t=1}^{r_a} \mu_{ta}^L\right)|\psi^{(L)}(A)\rangle$, with an integer $r_a$ and complex nonzero $\mu_{ta}$. Hence we obtain $\lambda_a^{(L)}=\sum_{t=1}^{r_a} \mu_{ta}^L$. Since there are finite number of 1D representations $\mathcal{L}_a\rightarrow \lambda_a^{(L)}|_{a\in\mathcal{B}}$ over $\mathbb{C}$ for the finite-dimensional algebra $\mathcal{R}$~\cite{Etingof2011}, $\mu_{ta}$ must be root of unity (see Lemma~\ref{lemma:period}). Hence there must be a common period $\ell$ for $\{\lambda_a^{(L)}\}_{a\in\mathcal{B}}$ as functions of $L$, and $\lambda_a^{(L)}=n_a$ (if we identify  $n_a=r_a$) when $L$ is an integer multiple of $\ell$. In other words, by blocking $\ell$ consecutive sites into a single effective site, the symmetry eigenvalues become independent of the system size.

An important class of MPO symmetries arises when they represent the fusion algebra of a unitary fusion category $\mathcal{C}$. In this case, there must be a unit of the algebra $\mathcal{R}$ represented by $O_I^{(L)}$, and each $O_a^{(L)}$ admits a dual $O_{\bar{a}}^{(L)}=[O_{a}^{(L)}]^{\dagger}$ such that $N_{ab}^I=\delta_{\bar{a},b}$. These properties are reflected in local tensors of MPOs, which capture the full categorical data of $\mathcal{C}$~\cite{Bultinck2017}. Accordingly, our anomaly-free condition is consistent with previously known results for categorical symmetry $\mathcal{C}$: 
a non-anomalous MPO symmetry admitting a symmetric normal MPS implies the existence of a module category for $\mathcal{C}$ with a single isomorphism class of simple objects (see \cite{GarreRubio2023} and Supplemental Material (SM) Sec.~\ref{sec:anomaly-free}~\cite{Supp}). This corresponds to the existence of a fiber functor, reflecting the anomaly-free condition for $\mathcal{C}$~\cite{Thorngren2024}. Furthermore, a direct consequence of Lemma~\ref{lemma:integral}, proven in the End Matter, provides a general instance when the MPO symmetry is anomalous:
\begin{thm}
An MPO symmetry representing a fusion algebra $\mathcal{R}$ (over $\mathbb{C}$) of a unitary fusion category $\mathcal{C}$ is anomalous if the quantum dimension of some simple object in $\mathcal{C}$ is a non-integer.
\label{thm:non-integer}
\end{thm}

\emph{Mixed states with strong anomalous symmetries}.-- Now let us generalize the discussion of MPO symmetries to mixed states. We consider a TI mixed state described by an MPDO $\rho^{(L)}(\mathcal{X},M)$~\cite{Verstraete2004,Zwolak2004}:
\begin{equation}
\begin{split}
\rho^{(L)}(\mathcal{X},M)=\sum_{\{i\},\{j\}}&\text{tr}[\mathcal{X}M^{i_1j_1}M^{i_2j_2}...M^{i_L j_L}]\\
&|i_1 i_2...i_L\rangle\langle j_1 j_2...j_L|,
\end{split}
\end{equation}
where $\mathcal{X}$ commutes with all $M^{ij}$. We are particularly interested in the case when the MPDO is strongly symmetric under $\{O_a^{(L)}\}_{a\in \mathcal{B}}$, in the sense that $O_a^{(L)}\rho^{(L)}(\mathcal{X},M)=\lambda_a^{(L)} \rho^{(L)}(\mathcal{X},M)$ for all $L$. The strong symmetry becomes a special case of the weak symmetry, defined via the condition $[O_a^{(L)},\rho^{(L)}]=0$ for all ${a\in\mathcal{B}}$ and all $L$, provided that a conjugate symmetry operator $O_{\bar{a}}^{(L)}=O_a^{(L)\dagger}$ exists for all $a\in\mathcal{B}$. 
The strong symmetry means that any state $|\phi^{(L)}\rangle$ within the support of $\rho^{(L)}(\mathcal{X},M)$ is symmetric: $O_a^{(L)}|\phi^{(L)}\rangle=\lambda_a^{(L)}|\phi^{(L)}\rangle$. It is straightforward to prove: 
\begin{lemma}
    For an MPDO $\rho^{(L)}(\mathcal{X},M)$ with a strong MPO symmetry $\{O_a^{(L)}\}$,  we have $(O_a^{(L)}\otimes \mathbb{1}_{\text{anc}}) |\varPsi^{(L)}\rangle=\lambda_a^{(L)}|\varPsi^{(L)}\rangle$ for any of its purifications $|\varPsi^{(L)}\rangle$, with $\mathbb{1}_{\text{anc}}$ the identity operator acting on the ancillas~\cite{Lessa2025a}.
    \label{prop:purification}
\end{lemma}
\noindent Now we prove the following no-go theorem:
\begin{thm}
Any MPDO $\rho^{(L)}(\mathcal{X},M)$ with a strong anomalous MPO symmetry as defined above cannot be prepared from TI normal MPSs by a local quantum channel, which adopts a Stinespring's dilation~\cite{Stinespring1955} to a TI matrix product unitary $W^{(L)}$, for all $L$. Here
\begin{equation}
W^{(L)}=
\begin{tikzpicture}[scale=0.45]
\draw[thick] (-0.3,0.5) arc (90:270:0.15);
\draw[thick] (0,0.5) -- (-0.3,0.5);
\QCA{(0,0)};
\draw[thick] (1,0.5)--(1.3,0.5);
\QCA{(1.3,0)};
\draw[thick] (2.3,0.5)--(2.6,0.5);
\QCA{(2.6,0)};
\draw[thick] (3.6,0.5)--(3.9,0.5);
\node at (4.3,0.5) {...};
\draw[thick] (4.7,0.5)--(5.0,0.5);
\QCA{(5.0,0)}
\draw[thick] (6.0,0.5)--(6.3,0.5);
\draw[thick] (6.3,0.5) arc (90:-90:0.15);
\end{tikzpicture}
\ ,
\end{equation}
where the shaded boxes represent $L$-independent identical tensors of finite dimension and black/blue legs denote physical/ancilla qudits.
\label{thm:nogo}
\end{thm}

\noindent The considered channel is a TI causal channel in dQC (defined in \cite{Piroli2020}), which can capture a TI finite-depth channel circuit. Similar TI ansatz was used for studying stability of mixed state SPT phases~\cite{deGroot2022,Xue2024}.

\emph{Proof:} Consider a TI MPDO generated from a trivial short-range correlated TI normal MPS $|\psi_0^{(L)}\rangle$ by the local quantum channel defined above:
\begin{equation}
\tilde{\rho}^{(L)}=\text{tr}_{\text{anc}}[W^{(L)}|\psi_0^{(L)}\rangle \langle\psi_0^{(L)}|\otimes(|0\rangle\langle 0|)^{\otimes L} W^{(L)\dagger}],
\end{equation}
where $\text{tr}_{\text{anc}}$ performs the trace over the ancillas. Let us assume $\tilde{\rho}^{(L)}$ has a strong MPO symmetry $\{O_a^{(L)}\}_{a\in\mathcal{B}}$, and prove it is anomaly-free, so that Theorem \ref{thm:nogo} holds. From Lemma \ref{prop:purification}, the pure TI MPS state 
\begin{equation}
    |\varPsi^{(L)}\rangle=W^{(L)}\left(|\psi_0^{(L)}\rangle\otimes |0\rangle^{\otimes L}\right)
\end{equation}
must be symmetric under $O_a^{(L)}\otimes\mathbb{1}_{\text{anc}}$, i.e., $(O_a^{(L)}\otimes \mathbb{1}_{\text{anc}})|\varPsi^{(L)}\rangle=\lambda_a^{(L)} |\varPsi^{(L)}\rangle$. Furthermore, $|\varPsi^{(L)}\rangle$ must be a normal MPS from properties of the matrix product unitary as a quantum cellular automaton~\cite{Cirac2017mpu}: $|\psi_0^{(L)}\rangle\otimes |0\rangle^{\otimes L}$ is the unique ground state of a strictly local parent Hamiltonian $H^{(L)}$. Hence $|\varPsi^{(L)}\rangle$ is the unique ground state of strictly local Hamiltonian $W^{(L)}H^{(L)} W^{(L)\dagger}$ for all $L$, which cannot be a non-normal MPS. Therefore the MPO symmetry must be anomaly-free. 

\emph{Nontrivial MPDO RFPs.--}
Tensor networks provide a powerful framework for understanding and classifying quantum phases of matter~\cite{Schuch2011,Cirac2021}. In 1D, mixed states described  by MPDOs adopt a natural real-space renormalization procedure~\cite{Cirac2017} similar to the pure-state MPSs~\cite{Verstraete2005} and the corresponding renormalization fixed points (RFPs) serve as representative states characterizing distinct quantum phases. For pure states, the MPS RFPs classify gapped phases with and without symmetries~\cite{Chen2011,Schuch2011}. In the mixed-state cases, although RFPs of MPDOs can be systematically described~\cite{Cirac2017}, their preparation complexity, which is crucial for defining topological phases for mixed states~\cite{Coser2019,Sang2024}, remains unclear.

Here we show that a general class of RFPs exhibit strong anomalous MPO symmetries and hence have fundamental obstructions to efficient preparation with local quantum channels. The local tensors can be constructed~\cite{Molnar2022} from a biconnected $C^*$-weak Hopf algebra (WHA) $\mathcal{A}$~\cite{Etingof2015} such that $\text{Rep}(\mathcal{A}^*)\cong \mathcal{C}$, where $\mathcal{C}$ is the corresponding anomalous categorical symmetry.  An explicit construction of $\mathcal{A}$ for arbitrary $\mathcal{C}$ can be found in SM Sec.~\ref{subsec:reconstruction}~\cite{Supp}. From these local tensors, we will explicitly construct nontrivial MPDO RFPs with strong anomalous MPO symmetry, which, due to our no-go theorem, are not efficiently preparable. In contrast, an anomaly-free unitary fusion category symmetry that admits a fiber functor corresponds to the case where $\mathcal{A}$ is a $C^*$-Hopf algebra~\cite{Etingof2015}, ensuring the efficient preparation of the constructed MPDO RFPs~\cite{Ruiz2024}. 

Let us briefly review necessary properties of a $C^*$-WHA $\mathcal{A}$. It is equipped with multiplication $\cdot$ and comultiplication $\Delta$:
\begin{equation}
x\cdot y=\sum_{z\in B_{\mathcal{A}}}\Lambda_{xy}^z z,\quad \Delta(z)=\sum_{x,y\in B_{\mathcal{A}}}\Gamma^z_{xy}x\otimes y,
\end{equation}
where $B_{\mathcal{A}}$ is the basis of $\mathcal{A}$, $x,y,z\in B_{\mathcal{A}}$, and the comultiplication is multiplicative, i.e., 
\begin{equation}
\Delta(x\cdot y)=\Delta(x)\cdot\Delta(y).
\label{eq:multiplicative}
\end{equation}
The full definition is in SM Sec.~\ref{subsec:definition}~\cite{Supp}, which involves additional properties including the unit and counit axioms, the antipode, and the $C^*$-structure. The dual algebra $\mathcal{A}^*$ is also a $C^*$-WHA with multiplication and comultiplication defined through the dual basis $\{\delta_x\}_{x\in B_{\mathcal{A}}}$ satisfying $\delta_x(y)=\delta_{xy}$ for $x,y\in B_{\mathcal{A}}$:
\begin{equation}
\begin{split}
&\delta_x\cdot \delta_y=\sum_{z\in B_{\mathcal{A}}}\Gamma_{xy}^z \delta_z, \quad \Delta(\delta_z)=\sum_{x,y\in B_{\mathcal{A}}}\Lambda^z_{xy}\delta_x\otimes \delta_y.\\
\end{split}
\end{equation}
The representations of $\mathcal{A}$ and $\mathcal{A}^*$ admit monoidal products: for representations $\Phi_{1,2}$ of $\mathcal{A}$ and $\Psi_{1,2}$ of $\mathcal{A}^*$, their composites $\Phi_1\boxtimes\Phi_2:=(\Phi_1 \otimes \Phi_2) \circ \Delta$ and $\Psi_1\boxtimes\Psi_2:=(\Psi_1 \otimes \Psi_2) \circ \Delta$ are again representations, as guaranteed by Eq.~\eqref{eq:multiplicative}. In fact, the representation categories of $\mathcal{A}$ or $\mathcal{A}^*$ form unitary fusion categories, whose simple objects are irreducible representations (irreps) and whose morphisms are intertwiners.

A nontrivial MPO algebra can be constructed from the following tensor~\cite{Molnar2022}:
\begin{equation}
T=\sum_{x\in B_{\mathcal{A}}}\Phi(x)\otimes \Psi(\delta_x)\equiv
\label{eqn:black-tensor-constr}
\begin{tikzpicture}[scale=0.6]
    \draw[thick,Red] (0,0)--(1,0); 
    \filldraw[black, ultra thin] (0.5,0) circle[radius=0.15];
    \draw[thick] (0.5,-0.5)--(0.5,0.5);
\end{tikzpicture}\ ,
\end{equation}
where $\Phi$ is a faithful *-representation of $\mathcal{A}$ in the physical space (black), and $\Psi$ is a faithful representation of  $\mathcal{A}^{*}$ in the virtual space (red). 
Let us choose $\Phi$ as $\Phi(x) = \oplus_\alpha \Phi_\alpha(x)$, where $\alpha$ labels all inequivalent *-irreps of $\mathcal{A}$, and similarly choose $\Psi$ as $\Psi(\delta_x) = \oplus_a \Psi_a(\delta_x)$, where $a$ labels all inequivalent irreps of $\mathcal{A}^*$. For convenience, we assume $\Psi$ and $\Psi_a$ to be *-representations,  as a gauge choice that can be relaxed when appropriate. The contraction rule of the tensors is as follows:
\begin{equation}
\begin{split}
\label{eqn:black-tensor-constr}
\begin{tikzpicture}[scale=0.6]
    \draw[thick,Red] (0,0)--(1,0);
    \filldraw[black, ultra thin] (0.5,0) circle[radius=0.15];
    \draw[thick] (0.5,-0.5)--(0.5,0.5);
    \draw[thick,Red] (0.75,0)--(1.75,0);
    \filldraw[black, ultra thin] (1.25,0) circle[radius=0.15];
    \draw[thick] (1.25,-0.5)--(1.25,0.5);
\end{tikzpicture}
&=\sum_{x\in B_{\mathcal{A}}}\Phi^{\boxtimes 2}(x)\otimes \Psi(\delta_x),\\
\begin{tikzpicture}[scale=0.6]
    \draw[thick,Red] (0,0)--(1,0);
    \filldraw[black, ultra thin] (0.5,0) circle[radius=0.15];
    \draw[thick] (0.5,-0.5)--(0.5,0.5);
    \draw[thick,Red] (0,0.74)--(1,0.75);
    \filldraw[black, ultra thin] (0.5,0.75) circle[radius=0.15];
    \draw[thick] (0.5,0.25)--(0.5,1.25);
\end{tikzpicture}
&=\sum_{x\in B_{\mathcal{A}}}\Phi(x)\otimes \Psi^{\boxtimes 2}(\delta_x),
\end{split}
\end{equation}
where we use shorthand $\Phi^{\boxtimes 2}:=\Phi\boxtimes\Phi$, $\Psi^{\boxtimes 2}:=\Psi\boxtimes\Psi$, and similarly $\Phi^{\boxtimes n}$, $\Psi^{\boxtimes n}$ from coassotiativity. Intuitively, the contraction in vertical (or horizontal) direction corresponds to the monoidal product with $\Phi$ (or $\Psi$).

Now let us consider an MPO with a boundary matrix $\mathcal{X}$:
\begin{equation}
\begin{split}
O^{(L)}(\mathcal{X})
&\equiv \begin{tikzpicture}[scale=0.7]
    \draw[thick] (0.8, 0) -- (0.8, 0.6);
    \draw[Red,thick] (0.2, 0) -- (2.4, 0);
    \draw[thick] (1.8, 0) -- (1.8, 0.55);
    \draw[thick] (3.8,0) -- (3.8, 0.5);
    \draw[Red, thick] (3.2, 0) -- (3.7, 0);
    \node[anchor=center, text=Red] at (2.82, 0.15) {$\overset{L}\cdots$};
    \draw[Red,thick] (0.25, 0) arc[start angle=90, end angle=270, radius=0.25] -- (3.924, -0.5) arc[start angle=-90, end angle=90, radius=0.25];
    \filldraw[Red, fill=Red] 
    (0.35, 0) circle[radius=0.125];
    \draw[preaction = { draw, white, line width=3pt,  line cap = round }, thick] (0.8, -0.6) -- (0.8, 0.6);
    \draw[preaction = { draw, white, line width=3pt,  line cap = round }, thick] (1.8, -0.6) -- (1.8, 0.6);
    \draw[preaction = { draw, white, line width=3pt,  line cap = round }, thick] (3.8, -0.6) -- (3.8, 0.6);
    \node[anchor=center, font=\footnotesize, text=Red] at (0.35, 0.3) {$\mathcal{X}$};
    \filldraw[black, ultra thin] 
    (3.8, 0) circle[radius=0.125];
    \filldraw[black, ultra thin] 
    (0.8, 0) circle[radius=0.125];
    \filldraw[black, ultra thin] 
    (1.8, 0) circle[radius=0.125];
\end{tikzpicture}\\
&=\sum_{x\in B_{\mathcal{A}}} \text{tr}[\mathcal{X}\Psi(\delta_x)]
\Phi^{\boxtimes L}(x). 
\end{split}
\end{equation}
We require that $\mathcal{X}$ commutes with all $\Psi(\delta_x)$ to ensure translational invariance. By Schur's lemma, this implies that $\mathcal{X} = \oplus_a \mathcal{X}_a \mathbb{1}_a = \sum_a \mathcal{X}_a P_a$, where $P_a$ is the Hermitian projector onto irrep $\Psi_a$.  The MPO symmetry generators are defined by
\begin{equation}
    O_a^{(L)}:=O^{(L)}(P_a)=\sum_{x\in B_{\mathcal{A}}}\chi_a(\delta_x) \Phi^{\boxtimes L}(x),
\end{equation}
where $\chi_a$ is the character of irrep $a$ and $O_a^{(L)}$ can be constructed from tensor $T_a=\sum_{x\in B_{\mathcal{A}}}\Phi(x)\otimes\Psi_a(\delta_x)$. It is straightforward to show that $O_{a}^{(L)}O_{b}^{(L)}=\sum_c N^c_{ab}O_c^{(L)}$, where $N_{ab}^c$ are defined by the decomposition $\Psi_a\boxtimes\Psi_b\cong \oplus_{c}\mathbb{1}_{N_{ab}^c}\otimes\Psi_c$ and serve as the structure constants of a fusion algebra $\mathcal{R}$ associated with a unitary fusion category.  In particular, the representation theory of $C^*$-WHA ensures the existence of a unit irrep $I$, and a dual irrep $\bar{a}$ for each $a$ satisfying $O_a^{(L)}=O_{\bar{a}}^{(L)\dagger}$ (see SM Sec.~\ref{subsec:dual_rep}~\cite{Supp}). 

For any 1D irrep of the fusion algebra $\mathcal{R}$, we can construct a Hermitian operator projector $O^{(L)}(\Pi_m)$, which is the MPO representation of the corresponding central idempotent, with a strong MPO symmetry, i.e., 
\begin{equation}
O_a^{(L)}O^{(L)}(\Pi_m)=\lambda_{ma} O^{(L)}(\Pi_m).
\end{equation}
All 1D irreps of $\mathcal{R}$ arise as common eigenvectors of the regular representation matrices $N_a$, defined by $(N_a)_{cb}= N_{ab}^c$, with corresponding eigenvalues $\lambda_{ma}$. A canonical 1D irrep corresponds to the Frobenius--Perron eigenvector, for which $\lambda_{0a} = d_a$, where $d_a$ denotes the quantum dimension of the irrep $a$. The associated projector is given by $\Pi_0 = \frac{1}{\mathcal{D}^2} \sum_a d_a P_a$,  where $\mathcal{D}:= \sqrt{\sum_a d_a^2}$ is the total quantum dimension.

Topologically nontrivial MPDO RFPs with strong anomalous MPO symmetry can be constructed from $O^{(L)}(\Pi_m) $:
\begin{equation}
    \rho^{(L)}_{\text{RFP},m}=\frac{1}{\mathcal{N}_m}O^{(L)}(\Pi_m)\Omega^{\otimes L}, 
    \label{eq:RFP}
\end{equation}
where $\mathcal{N}_m$ is a $L$-independent normalization factor (see SM Sec.~\ref{subsec:normalization}~\cite{Supp}) and $\Omega$ is the dual of $\Pi_0$ in the physical space, i.e., $\Omega=\frac{1}{\mathcal{D}^2}\sum_{\alpha} \kronecker_{\alpha}Q_{\alpha}$. Here $\alpha$ labels an irrep of $\mathcal{A}$ with quantum dimension $\kronecker_{\alpha}$, and $Q_{\alpha}$ is the Hermitian projector from $\Phi$ onto irrep $\alpha$. The total quantum dimensions for irreps of $\mathcal{A}$ and $\mathcal{A}^*$ coincide for biconnected WHA ~\cite{Ruiz2024,Etingof2015}, and thus share the same notation. The construction yields valid density operators since $\Omega^{\otimes L}$ and $O^{(L)}(\Pi_m)$ are commuting positive operators. The construction and properties of the corresponding MPDO tensor $M=\begin{tikzpicture}[baseline={([yshift=-1.2ex]current bounding box.center)},scale=0.5]
    \draw[thick,Red] (0,0)--(1,0);
    \filldraw[black, ultra thin] (0.5,0) circle[radius=0.15];
   \draw[thick] (0.5,-0.5)--(0.5,0.75);
    \filldraw[black, ultra thin] (0.4,0.4) rectangle (0.6,0.6);
    \node at (0.1,0.5) {\small $\Omega$};
\end{tikzpicture}$ were studied in \cite{Ruiz2024}. Here we focus on the case with strong symmetry, realized by a subset of boundary matrices $\mathcal{X}=\Pi_m$.

It is worth noting that these MPDO RFPs are locally indistinguishable from a weakly symmetric state $\rho_0^{(L)}\sim O_I^{(L)} \Omega^{\otimes L}$ satisfying $[O_a^{(L)},\rho_0^{(L)}]=0$: Although $\rho^{(L)}_{\text{RFP},m}$ contains linear combinations of $O_a^{(L)}\Omega^{\otimes L}$, one can prove (see SM Sec.~\ref{subsec:trace-one-site}~\cite{Supp}) that its correlation length is zero, and after tracing out any site, the only nonvanishing term is $O_I^{(L)} \Omega^{\otimes L}$. This behavior mirrors the phenomenon of strong-to-weak spontaneous symmetry breaking~\cite{Lee2023,Ma2023b,Sala2024,Lessa2025swssb,Gu2024} discussed in the unitary symmetry case.

\emph{Preparation with measurements and feedforward.--}
A natural question is whether the constructed MPDO RFPs with strong anomalous symmetry can be efficiently prepared via quantum circuit enhanced by local measurements and (nonlocal) classical feedforward, a scheme capable to generate long-range entanglement (see ~\cite{Briegel2001,Raussendorf2005} and recent developments, e.g., ~\cite{Piroli2021,Tantivasadakarn2024,Verresen2021, Bravyi2022, Lu2022,Tantivasadakarn2023hierarchy,Tantivasadakarn2023shortest}). We answer this affirmatively by showing that $\rho^{(L)}_{\text{RFP,m}}$ is locally purifiable to a fixed-point MPS known to be preparable via finite-depth circuit with measurements and feedforward. This becomes evident upon rewriting the MPDO in the form of a locally purifiable density operator:
\begin{equation}
\rho^{(L)}_{\text{RFP},m}=\frac{1}{\mathcal{N}_m} \begin{tikzpicture}[scale=0.9]
    \filldraw[black, ultra thin] (0.8-0.1*2/3,0.4-0.1*2/3) rectangle (0.8+0.1*2/3,0.4+0.1*2/3);
    \node at (1.2,0.4) {\footnotesize $\sqrt{\Omega}$};
    \filldraw[black, ultra thin] (1.8-0.1*2/3,0.4-0.1*2/3) rectangle (1.8+0.1*2/3,0.4+0.1*2/3);
    \node at (2.2,0.4) {\footnotesize $\sqrt{\Omega}$};
    \filldraw[black, ultra thin] (3.8-0.1*2/3,0.4-0.1*2/3) rectangle (3.8+0.1*2/3,0.4+0.1*2/3);
     \node at (4.2,0.4) {\footnotesize $\sqrt{\Omega}$};
    \draw[thick] (0.8, 0) -- (0.8, 0.6);
    \draw[Red,thick] (0.2, 0) -- (2.4, 0);
    \draw[thick] (1.8, 0) -- (1.8, 0.6);
    \draw[thick] (3.8,0) -- (3.8, 0.6);
    \draw[Red, thick] (3.2, 0) -- (3.7, 0);
    \node[anchor=center, text=Red] at (2.82, 0.15) {$\overset{L}\cdots$};
    \draw[Red,thick] (0.25, 0)--(-0.1, 0) arc[start angle=90, end angle=270, radius=0.1];
    \draw[Red,thick] (3.8, 0)--(4.15, 0) arc[start angle=90, end angle=-90, radius=0.1];

    \filldraw[black, ultra thin] (0.8-0.1*2/3,-1-0.1*2/3) rectangle (0.8+0.1*2/3,-1+0.1*2/3);
    \node at (1.2,-1) {\footnotesize $\sqrt{\Omega}$};
    \filldraw[black, ultra thin] (1.8-0.1*2/3,-1-0.1*2/3) rectangle (1.8+0.1*2/3,-1+0.1*2/3);
    \node at (2.2,-1) {\footnotesize $\sqrt{\Omega}$};
    \filldraw[black, ultra thin] (3.8-0.1*2/3,-1-0.1*2/3) rectangle (3.8+0.1*2/3,-1+0.1*2/3);
     \node at (4.2,-1) {\footnotesize $\sqrt{\Omega}$};
    \filldraw[Red, fill=Red] 
    (0.35, 0) circle[radius=0.1];
    \draw[thick] (0.8, -0.6) -- (0.8, 0.6);
    \draw[thick] (1.8, -0.6) -- (1.8, 0.6);
    \draw[thick] (3.8, -0.6) -- (3.8, 0.6);
    \node[anchor=center, font=\footnotesize, text=Red] at (0.25, 0.4) {$\Pi_m$};
        \draw[thick] (0.8, -0.6) -- (0.8, 0);
    \draw[Red,thick] (0.2, -0.6) -- (2.4, -0.6);
    \draw[thick] (1.8, -0.6) -- (1.8, 0);
    \draw[thick] (3.8, -0.6) -- (3.8, 0);
    \draw[Red, thick] (3.2, -0.6) -- (3.7, -0.6);
    \node[anchor=center, text=Red] at (2.82, -0.45) {$\overset{L}\cdots$};
    \draw[Red,thick] (0.25,-0.6)--(-0.1,-0.6) arc[start angle=90, end angle=270, radius=0.1];
    \draw[Red,thick] (3.8, -0.6)--(4.15, -0.6) arc[start angle=90, end angle=-90, radius=0.1];
    \filldraw[Red, fill=Red] 
    (0.35, -0.6) circle[radius=0.1];
    \draw[thick] (0.8, -1.2) -- (0.8, 0);
    \draw[thick] (1.8, -1.2) -- (1.8, 0);
    \draw[thick] (3.8, -1.2) -- (3.8, 0);
    \node[anchor=center, font=\footnotesize, text=Red] at (0.25, -1) {$\Pi^*_m$};
    \draw[black, thick, fill=white] 
    (3.8, -0.6) circle[radius=0.1];
    \draw[black, thick,fill=white] 
    (0.8, -0.6) circle[radius=0.1];
    \draw[black, thick,fill=white] 
    (1.8, -0.6) circle[radius=0.1];
    \filldraw[black, ultra thin] 
    (3.8, 0) circle[radius=0.1];
    \filldraw[black, ultra thin] 
    (0.8, 0) circle[radius=0.1];
    \filldraw[black, thick, ultra thin] 
    (1.8, 0) circle[radius=0.1];
   \end{tikzpicture}, 
\end{equation}
where we have used property of $O^{(L)}(\Pi_m)$ being a Hermitian projector. The periodic boundary condition is not explicitly drawn. The upper half of the diagram represents Hermitian operator $O^{(L)}(\Pi_m) \sqrt{\Omega}^{\otimes L}$, and the lower half its Hermitian conjugation, applied at the level of each local tensor.
This gives a purification to an MPS, i.e., $\frac{1}{\sqrt{\mathcal{N}_m}}O^{(L)}(\Pi_m) \sqrt{\Omega}^{\otimes L}$ in the doubled Hilbert space, which can be proven to be an MPS fixed point known to be preparable by a finite-depth circuit assisted with measurements and feedforward~\cite{Piroli2021} (see End Matter for a detailed proof and an explicit preparation protocol).

\emph{Example of Fibonacci fusion category $\mathcal{C}_{\text{Fib}}$:} 
Let us discuss an example of MPDO RFPs constructed from Fibonacci fusion category, which exhibits anomalous MPO symmetries due to the presence of a non-integer quantum dimension. More examples, such as $\text{Vec}_{G}^{\omega}$ for any finite group $G$ with nontrivial 3-cocycle, can be found in SM Sec.~\ref{sec:group_cohomology}~\cite{Supp}. One can reconstruct a $C^*$-WHA $\mathcal{A}_{\text{Fib}}$ satisfying $\mathcal{A}_{\text{Fib}}\cong \mathcal{A}^*_{\text{Fib}}\cong \mathcal{M}_2(\mathbb{C})\oplus \mathcal{M}_3(\mathbb{C})$~\cite{Bohm1996,Ruiz2024}, where $\mathcal{M}_n(\mathbb{C})$ is the algebra of complex $n\times n$ matrices. The basis consists of $e_{\alpha,ij}$ with $i,j\in\{1,2\}$ for $\alpha=I$ and $i,j\in\{1,2,3\}$ for $\alpha=\tau$, which can be represented faithfully as $\Phi(e_{\alpha,ij})=|\alpha,i\rangle \langle \alpha,j|$. The corresponding dual basis element $\tilde{e}_{\alpha,ij}$ is represented by $\Psi(\tilde{e}_{\alpha,ij})=\sum_{a,k,l}R_{a,kl}^{\alpha,ij}|a,k)(a,l|$, where the components of matrix $R$ can be found in SM Sec.~\ref{sec:Fibonacci}~\cite{Supp}. The vectors $|\alpha,i\rangle$ and $|a,k)$ span the irrep spaces labeled by $I$ and $\tau$, with $i,k \in \{1,2\}$ when $a,\alpha = I$ and $i,k \in \{1,2,3\}$ when $a,\alpha = \tau$.  The operator $\Omega$ can be explicitly constructed from their quantum dimensions:
\begin{equation}
\Omega=\frac{2}{5+\sqrt{5}}Q_I+\frac{1}{\sqrt{5}}Q_{\tau}.
\end{equation} 
There are two boundary matrices, $\Pi_0$ and $\Pi_1$, yielding locally indistinguishable MPDO RFPs with strong MPO symmetry (detailed in SM Sec.~\ref{sec:Fibonacci}~\cite{Supp}):
\begin{equation}
\Pi_0=\frac{2}{5+\sqrt{5}}P_I+\frac{1}{\sqrt{5}}P_{\tau}, \quad \Pi_1=\frac{2}{5-\sqrt{5}}P_I-\frac{1}{\sqrt{5}}P_{\tau}.
\label{eq:Fib-Pi}
\end{equation}
The possibility of nontrivial MPDO RFPs constructed from $\mathcal{A}_{\text{Fib}}$ was conjectured in \cite{Ruiz2024}. Here, the two constructed MPDO RFPs exhibit strong anomalous MPO symmetry, which provides the obstruction to efficient preparation from Theorem \ref{thm:nogo}. Furthermore, to guide future experimental studies, we provide details to prepare these states through measurement and feedforward in the End Matter.

\emph{Conclusion and outlook.--}  In conclusion, we have proven that MPDOs with strong anomalous MPO symmetries, encompassing both unitary and non-invertible symmetries, cannot be prepared from trivial states (normal MPSs) via any TI finite-depth local quantum channel. We constructed a general family of MPDO RFPs that realize such symmetries. While these fixed points lie beyond the efficiently preparable class, they remain accessible through finite-depth quantum circuits enhanced by measurements and feedforward. Our results reveal a fundamental obstruction to state preparation posed by anomalous generalized symmetries for mixed states. Future directions include identifying physical signatures of these nontrivial RFPs, extending classification schemes beyond anomaly-based obstructions, and exploring generalizations to higher dimensions within the tensor network framework.

\begin{acknowledgments}
I thank Ansi Bai, Meng Cheng, Ignacio Cirac, Zongping Gong, Ze-Min Huang, András Molnár, David Pérez-García, Alberto Ruiz-de-Alarcón, and Zhiyuan Wang for helpful discussions. This work is supported by the Alexander von Humboldt Foundation.
\end{acknowledgments}
\bibliography{refs}
\section*{End Matter}
\section{Additional details for Lemma 1}
\label{sec:Lemma1}
In the proof of Lemma~\ref{lemma:integral}, we used the following Lemma:
\begin{lemma}
Let $\{z_1,z_2,..., z_s\}$ be $s$ nonzero complex numbers. Define $\mathcal{F}(L):=\sum_{t=1}^s z_t^L$ as a complex function of positive integer $L$. If the image of $\mathcal{F}$ is a finite set, then each $z_t$ is a root of unity and $\mathcal{F}$ is a periodic function.
\label{lemma:period}
\end{lemma}

\emph{Proof}:  The sequence $\mathcal{F}(L)$ satisfies a linear recurrence relation
\begin{equation}
\sum_{t=0}^{s} C_t \mathcal{F}(L+s-t)=0,\quad C_t:=(-1)^t  \mathcal{E}_t(z_1,z_2,..., z_s),
\label{eq:recursion}
\end{equation}
where $\mathcal{E}_t(z_1,z_2,..., z_s)$ is the $t$-th elementary symmetric polynomial, i,.e.,
\begin{equation}
\mathcal{E}_t(z_1,z_2,..., z_s):=\sum_{1\le n_1<n_2...<n_t\le s }z_{n_1} z_{n_2}...z_{n_t},
\end{equation}
and $\mathcal{E}_0(z_1,z_2,..., z_s):=1$. Indeed, the characteristic polynomial
\begin{equation}
\mathcal{P}(z)=\prod_{t=1}^s (z-z_t)=\sum_{t=0} z^{s-t}(-1)^t\mathcal{E}_{t}(z_1,z_2,...,z_s).
\end{equation}
implies
\begin{equation}
\sum_{t'=1}^s z_{t'}^{L} \sum_{t} z_{t'}^{s-t}(-1)^t\mathcal{E}_{t}(z_1,z_2,...,z_s)=\sum_{t'=1}^s z_{t'}^{L} \mathcal{P}(z_{t'})=0,
\end{equation}
which yields Eq.~\eqref{eq:recursion}. Thus $\mathcal{F}(L+s)$ and $\mathcal{F}(L-1)$ are uniquely determined by $\{\mathcal{F}(L),\dots,\mathcal{F}(L+s-1)\}$.

Define 
\begin{equation}
\mathcal{U}(L):=(\mathcal{F}(L),\mathcal{F}(L+1),...,\mathcal{F}(L+s-1)).
\end{equation}
Since $\mathcal{F}$ has finite image, so does $\mathcal{U}$. By the pigeonhole principle, there exist $L_0$ and $\ell$ such that $\mathcal{U}(L_0)=\mathcal{U}(L_0+\ell)$. The recurrence then implies
\begin{equation}
\mathcal{F}(L+\ell)-\mathcal{F}(L)=\sum_{t}^s[(z_t^\ell-1)z_t] z_t^{L-1}=0
\end{equation}
for all $L$. Let $\mathcal{S}:=\{u_1,u_2,...,u_{s_0}\}$ be the set of distinct values among $\{z_t\}$, with $s_0\le s$. For each $u_j\in\mathcal{S}$, define the vector 
\begin{equation}
w_{j}:=(1, u_j^1,..., u_j^{s_0-1})
\end{equation}
These vectors are linearly independent by the nonvanishing of the Vandermonde determinant. Now consider $\mathcal{F}(L+\ell)-\mathcal{F}(L)=0$ for $L=1,2,..., s_0$, we have 
\begin{equation}
\sum_{j=1}^{s_0}  \left[\sum_{t|z_t=u_{j}}(z_t^{\ell}-1)z_t\right]w_{j}=0
\end{equation}
By linear independence of $\{w_{j}\}$, the coefficient $\left[\sum_{t|z_t=u_{j}}(z_t^{\ell}-1)z_t\right]$ must vanish for all $j=1,...,s_0$, which is $(u_{j}^{\ell}-1)u_{j}$ times the number of $z_t$ that equals to $u_{j}$. Since $u_{j}$ is nonzero, we must havet $u_{j}^{\ell}=1$, meaning that all of $u_{j}$, hence all of $z_t$,  are roots of unity.

\section{Proof of Theorem 1}
\label{sec:Theorem1}
We prove the equivalent statement that for any non-anomalous MPO symmetry $\{O_a^{(L)}\}_{a\in\mathcal{B}}$ representing a fusion algebra $\mathcal{R}$ of a unitary fusion category $\mathcal{C}$, the Frobenius-Perron dimensions of basis elements $\{\mathcal{L}_a\}$ (identical to the quantum dimensions of the associated simple objects in $\mathcal{C}$) must be integers. For such a non-anomalous MPO symmetry,  there exists a symmetric normal MPS under $\{O_a^{(L)}\otimes\mathbb{1}_r^{\otimes L}\}_{a\in\mathcal{B}}$ with some finite $r$. Then from Lemma 1 in the main text, there exists non-negative integers $n_a$ such that $\mathcal{L}_a\rightarrow n_a $ for $a\in\mathcal{B}$ forms 1D irrep of $\mathcal{R}$ over $\mathbb{C}$.  
According to Proposition 3.3.6 in \cite{Etingof2015}, $n_a$ must be the Frobenius-Perron dimension of $\mathcal{L}_a$ hence the theorem is proven.

\section{$\frac{1}{\sqrt{\mathcal{N}_m}} O^{(L)}(\Pi_m)\sqrt{\Omega}^{\otimes L}$ corresponds to an MPS fixed point in doubled Hilbert space}
\label{sec:MPS_fixed_point}
It suffices to prove that (i) $O^{(L)}_a\sqrt{\Omega}^{\otimes L}|_{a\in\mathcal{B}}$ in the doubled Hilbert space correspond to normal MPS fixed points and (ii) they are locally orthogonal to each other according to Definition 3.5 in \cite{Cirac2017}. Let us keep the notation for $T_a:=\sum_{x\in B_{\mathcal{A}}}\Phi(x)\otimes \Psi_a(\delta_x) \equiv \begin{tikzpicture}[scale=0.5]
    \draw[thick,Red] (0,0)--(1,0);
    \filldraw[black, ultra thin] (0.5,0) circle[radius=0.15];
    \draw[thick] (0.5,-0.5)--(0.5,0.5);
    \node [anchor=center, font=\footnotesize, text=Red]  at (-0.3,0) {\small $a$};
\end{tikzpicture}\ $ which generates $O_a^{(L)}$. Since $[O_a^{(L)}]^{\dagger}=O_{\bar{a}}^{(L)}$ for all $L$,  from the fundamental theorem~\cite{Cirac2017}, taking the Hermitian conjugation of $T_a$ in the physical space must generate a tensor that differs from $T_{\bar{a}}$ by a gauge transformation, i.e., 
\begin{equation}
\begin{split}
\begin{tikzpicture}[scale=0.5]
    \draw[thick,Red] (-0.2,0)--(1.2,0);
    \draw[thick] (0.5,-0.7)--(0.5,0.7);
    \node [anchor=center, font=\footnotesize, text=Red]  at (-0.5,0) {\small $a$};
    \draw[black, thick, fill=white] (0.5,0) circle[radius=0.15];
\end{tikzpicture}&:=\sum_{x\in B_{\mathcal{A}}} \Phi^{\dagger}(x)\otimes \overline{\Psi_{a}(\delta_x)}\\
&=\sum_{x\in B_{\mathcal{A}}} \Phi(x)\otimes \tilde{\Psi}_{\bar{a}}(\delta_x),
\end{split}
\end{equation}
where despite that $\tilde{\Psi}_{\bar{a}}$ may not be a $*$-irrep, $\tilde{\Psi}_{\bar{a}}\cong \Psi_{\bar{a}}$ as a representation, i.e., $\tilde{\Psi}_{\bar{a}}(\delta_x)=\tilde{V}_{\bar{a}}^{-1}\Psi_{\bar{a}}(\delta_x)\tilde{V}_{\bar{a}}$ for all $x\in B_{\mathcal{A}}$. Now consider the transfer matrix for the overlap between normal MPSs representing $O^{(L)}_a\sqrt{\Omega}^{\otimes L}$ and $O^{(L)}_b\sqrt{\Omega}^{\otimes L}$, respectively, in the doubled Hilbert space:
\begin{equation}
E_{ab}=\begin{tikzpicture}[scale=0.6]
    \draw[black,thick] (0.5, 1.3) arc[start angle=180, end angle=0, radius=0.25]--(1,-1) arc[start angle=0, end angle=-180, radius =0.25];
    \filldraw[black, ultra thin] (0.4,0.9) rectangle (0.6,1.1);
    \node at (0.05,1) {\small $\Omega$};
    \draw[preaction = { draw, white, line width=3pt,  line cap = round }, thick,Red] (-0.2,0.5)--(1.2,0.5);
    \filldraw[black, ultra thin] (0.5,0.5) circle[radius=0.15];
    \draw[thick] (0.5,0)--(0.5,1);
    \node [anchor=center, font=\footnotesize, text=Red]  at (-0.5,0.5) {\small $a$};
    \draw[preaction = { draw, white, line width=3pt,  line cap = round }, thick,Red] (-0.2,-0.5)--(1.2,-0.5);
    \draw[thick] (0.5,-1)--(0.5,0);
    \node [anchor=center, font=\footnotesize, text=Red]  at (-0.5,-0.5) {\small $b$};
    \draw[black, thick, fill=white] (0.5,-0.5) circle[radius=0.15];
    \draw[black, thick] (0.5,1)--(0.5,1.3);
\end{tikzpicture}=\sum_{x\in B_{\mathcal{A}}}\text{tr}[\Omega \Phi(x)][\Psi_a\boxtimes \tilde{\Psi}_{\bar{b}}](\delta_x).
\end{equation}
The spectrum of $E_{ab}$ coincides with that of
\begin{equation}
\begin{split}
\tilde{E}_{ab} &=\bigoplus_{c}\sum_{x\in B_{\mathcal{A}}}\text{tr}[\Omega \Phi(x)][ \mathbb{1}_{N_{a\bar{b}}^c} \otimes\Psi_c(\delta_x) ]\\
&=\bigoplus_{c}[\mathbb{1}_{N_{a\bar{b}}^c}\otimes \Psi_c(\theta) ]\
\end{split}
\end{equation}
where we have used $\Psi_a\boxtimes\tilde{\Psi}_{\bar{b}}\cong \bigoplus_{c}\mathbb{1}_{N_{a\bar{b}}^c}\otimes \Psi_c(\delta_x) $ and $\theta:=\sum_{x\in B_{\mathcal{A}}}\delta_x\text{tr}[\Omega\Phi(x)]$ is the regular canonical element in $\mathcal{A}^*$.  
It is known (see SM Sec.~\ref{subsec:trace-one-site}) that the only $\Psi_c(\theta)$ that is nonvanishing is $\Psi_I(\theta)$, which is a rank-1 projector. Hence  $\tilde{E}_{ab}=E_{ab}=0$ unless $a=b$, in which case the spectrum of the transfer matrix $E_{aa}$ has only one eigenvalue that is 1 while others are zero. Therefore $O^{(L)}_a\sqrt{\Omega}^{\otimes L}|_{a\in\mathcal{B}}$ correspond to normal MPS fixed points and they are locally orthogonal ($E_{ab}=0$), which implies that $\frac{1}{\sqrt{\mathcal{N}_m}} O^{(L)}(\Pi_m)\sqrt{\Omega}^{\otimes L}$ is an MPS fixed point in the doubled Hilbert space. 

\section{Preparation by measurement and feedforward}

By measurement and feedforward, we can prepare the purified state. Here we write the state explicitly:
\begin{equation}
\begin{split}
&\frac{1}{\sqrt{\mathcal{N}_m}}
    \sum_a \Pi_{ma}
    \begin{tikzpicture}[scale=0.9]
    \filldraw[black, ultra thin] (0.8-0.1*2/3,0.4-0.1*2/3) rectangle (0.8+0.1*2/3,0.4+0.1*2/3);
    \node at (0.4,0.4) {\footnotesize $\sqrt{\Omega}$};
    \filldraw[black, ultra thin] (1.8-0.1*2/3,0.4-0.1*2/3) rectangle (1.8+0.1*2/3,0.4+0.1*2/3);
    \node at (1.46,0.4) {\footnotesize $\sqrt{\Omega}$};
    \filldraw[black, ultra thin] (3.8-0.1*2/3,0.4-0.1*2/3) rectangle (3.8+0.1*2/3,0.4+0.1*2/3);
     \node at (3.4,0.4) {\footnotesize $\sqrt{\Omega}$};
    \draw[thick] (0.8, 0) -- (0.8, 0.6);
    \draw[Red,thick] (0.25, 0) -- (2.4, 0);
    \draw[thick] (1.8, 0) -- (1.8, 0.6);
    \draw[thick] (3.8,0) -- (3.8, 0.6);
    \draw[Red, thick] (3.2, 0) -- (4.2, 0);
    \node[anchor=center, text=Red] at (2.82, 0.15) {$\overset{L}\cdots$};
    \draw[Red,thick] (0.25, 0) arc[start angle=90, end angle=270, radius=0.1];
    \draw[Red,thick] (4.2, 0)--(4.55, 0) arc[start angle=90, end angle=-90, radius=0.1];
    \draw[thick] (0.8, -0.4) -- (0.8, 0.6);
    \draw[thick] (1.8, -0.4) -- (1.8, 0.6);
    \draw[thick] (3.8, -0.4) -- (3.8, 0.6);
    \draw[thick] (0.8, -0.4) -- (0.8, 0);
    \draw[thick] (1.8, -0.4) -- (1.8, 0);
    \draw[thick] (3.8, -0.4) -- (3.8, 0);
    \draw[thick,preaction={draw=white, line width=5pt}] (1.2,-0.4)--(1.2,0.2);
    \draw[thick] (1.2,0.2)--(1.2,0.6);
    \draw[thick,preaction={draw=white, line width=5pt}] (2.2,-0.4)--(2.2,0.6);
    \draw[thick,preaction={draw=white, line width=5pt}] (4.2,-0.4)--(4.2,0.6);
    \draw[thick] (0.8, -0.4) arc[start angle=180, end angle=360, radius=0.2];
    \draw[thick] (1.8, -0.4) arc[start angle=180, end angle=360, radius=0.2];
    \draw[thick] (3.8, -0.4) arc[start angle=180, end angle=360, radius=0.2];
    \node[Red] at (0.3,0.13) {\footnotesize $a$};
    \filldraw[black, ultra thin] 
    (3.8, 0) circle[radius=0.1];
    \filldraw[black, ultra thin] 
    (0.8, 0) circle[radius=0.1];
    \filldraw[black, thick, ultra thin] 
    (1.8, 0) circle[radius=0.1];
   \end{tikzpicture}\\
   =&\frac{1}{\sqrt{\mathcal{N}_m}}\sum_a \Pi_{ma}\begin{tikzpicture}[scale=0.9]
    \node[anchor=center, text=Red] at (3.05, 0.15) {$\overset{L}\cdots$};
    \draw[Red,thick] (0.25, 0) arc[start angle=90, end angle=270, radius=0.1];
    \draw[Red,thick] (4.2,0.5)--(4.2, 0)--(4.55, 0) arc[start angle=90, end angle=-90, radius=0.1];
    \node[Red] at (0.3,0.13) {\footnotesize $a$};
    \draw[Red,thick] (0.25, 0)--(0.8, 0) --(0.8,0.5);
    \draw[Black,thick] (0.8,0.5)--(0.8,0.8);
    \filldraw[Black,thick,line join=round] (0.42,0)--(0.55,0.1)--(0.68,0)--(0.55,-0.1)--cycle;
    \draw[Red,thick] (1, 0.5)--(1, 0) --(1.5,0)--(1.5,0.5);
    \draw[Black,thick] (1.5, 0.5)--(1.5,0.8);
    \draw[Black,thick] (1, 0.5)--(1,0.8);
    \filldraw[Gray,thick]  (0.7,0.2) rectangle (1.1,0.6);
    \filldraw[Black,thick,line join=round] (0.42+0.7,0)--(0.55+0.7,0.1)--(0.68+0.7,0)--(0.55+0.7,-0.1)--cycle;
    \draw[Red,thick] (1+0.7, 0.5)--(1+0.7, 0) --(1.5+0.7,0)--(1.5+0.7,0.5);
    \draw[Black,thick] (1+0.7, 0.5)--(1+0.7, 0.8);
    \draw[Black,thick] (1.5+0.7, 0.5)--(1.5+0.7, 0.8);
    \filldraw[Gray,thick]  (0.7+0.7,0.2) rectangle (1.1+0.7,0.6);
    \filldraw[Black,thick,line join=round] (0.42+1.4,0)--(0.55+1.4,0.1)--(0.68+1.4,0)--(0.55+1.4,-0.1)--cycle;
    \draw[Red,thick] (1+1.4, 0.5)--(1+1.4, 0) --(1.2+1.4,0);
    \draw[Black,thick] (1+1.4, 0.5)--(1+1.4, 0.8);
    \filldraw[Gray,thick]  (0.7+1.4,0.2) rectangle (1.1+1.4,0.6);
    \node[White] at (0.9,0.4) {\footnotesize $U$};
    \node[White] at (0.9+0.7,0.4) {\footnotesize $U$};
    \node[White] at (0.9+1.4,0.4) {\footnotesize $U$};
    \node[Black] at (0.55,-0.3) {\footnotesize $\Lambda_a$};
    \node[Black] at (0.55+0.7,-0.3) {\footnotesize $\Lambda_a$};
    \node[Black] at (0.55+1.4,-0.3) {\footnotesize $\Lambda_a$};
     \draw[Red,thick] (0.25+3.2, 0)--(0.8+3.2, 0) --(0.8+3.2,0.5);
    \draw[Black,thick] (0.8+3.2,0.5)--(0.8+3.2,0.8);
    \draw[Black,thick] (1+3.2,0.5)--(1+3.2,0.8);
    \filldraw[Black,thick,line join=round] (0.42+3.2,0)--(0.55+3.2,0.1)--(0.68+3.2,0)--(0.55+3.2,-0.1)--cycle;
    \filldraw[Gray,thick]  (0.7+3.2,0.2) rectangle (1.1+3.2,0.6);
    \node[White] at (0.9+3.2,0.4) {\footnotesize $U$};
    \node[Black] at (0.55+3.2,-0.3) {\footnotesize $\Lambda_a$};
   \end{tikzpicture}\\
   =&U^{\otimes L}\frac{1}{\sqrt{\mathcal{N}_m}}\sum_a \Pi_{ma}|\omega(\Lambda_a)\rangle^{\otimes L},
\end{split}
\end{equation}
where $\Pi_{ma}$ is defined via $\Pi_{m}=\sum_a \Pi_{ma}P_a$, and we have used the standard form of MPS RFPs, i.e., $U$ is an onsite unitary operator, $\Lambda_a$ is a Hermitian positive matrix, and $\{|\omega(\Lambda_a)\rangle\}:= \{\begin{tikzpicture}[baseline={([yshift=0.5ex]current bounding box.center)},scale=0.9]
    \draw[Red,thick] (1, 0.2)--(1, 0) --(1.5,0)--(1.5,0.2);
    \filldraw[Black,thick,line join=round] (0.42+0.7,0)--(0.55+0.7,0.1)--(0.68+0.7,0)--(0.55+0.7,-0.1)--cycle;
     \node[Black] at (0.55+0.7,-0.25) {\footnotesize $\Lambda_a$};
\end{tikzpicture}\}$ contains entangled pair between neighboring site that are mutually orthogonal. We order the $\mathcal{M}$ states $|\omega(\Lambda_a)\rangle$ as $i_a\in \{0,..., \mathcal{M}-1\}$, where $\mathcal{M}$ is the number of irreps $a$, and create a qudit $|i_a\rangle$. The state $|\omega(\Lambda_a)\rangle$ is obtained by applying on an isometry $W$ to the qudit $W|i_a\rangle=|\omega(\Lambda_a)\rangle$. In practice, we can choose $M$ reference states in the physical Hilbert space and implement the isometry $W$ as a unitary. 

Now the strategy to prepare the purified state is to prepare $\frac{1}{\sqrt{\mathcal{N}_m}}\sum_a \Pi_{ma}|i_a\rangle^{\otimes L}$ and then apply $U^{\otimes L}W^{\otimes L}$. The preparation of the state $\frac{1}{\sqrt{\mathcal{N}_m}}\sum_a \Pi_{ma}|i_a\rangle^{\otimes L}$ is similar to that of the Greenberger–Horne–Zeilinger (GHZ) state. This can be done in three steps:
\begin{enumerate}
\item We attach each qudit an ancilla except for the first one and prepare $\otimes_{n=1}^{N-1}|\Phi^+\rangle_{s_n,a_{n+1}}\otimes \frac{1}{\sqrt{\mathcal{N}_m}}\sum_a \Pi_{ma}|i_a\rangle_{s_N}$, where $|\Phi^+\rangle_{s_n,a_{n+1}}$ is the maximal entangled Bell pair between the qubit on $n$-th site and the ancilla on $(n+1)$-th site.
\item Apply a controlled modular substraction gate between each qubit and its ancilla (defined as SUB$_{n}(|i_n\rangle_{s_n}\otimes |j_n\rangle_{a_n})=|i_n\rangle_{s_n}\otimes |i_n\ominus j_n\rangle_{a_n}$ on site $n$), yielding 
\begin{equation}
\begin{split}
&\sum_{\{i_n\}}|i_1\rangle_{s_1}(\otimes_{n=2}^{N-1} |i_n\rangle_{s_n}\otimes |i_n\ominus i_{n-1}\rangle_{a_n})  \\
&\otimes\frac{1}{\sqrt{\mathcal{N}_m}}\sum_a (\Pi_{ma} |i_a\rangle_{s_N}\otimes |i_a\ominus i_{N-1}\rangle_{a_N}),
\end{split}
\end{equation}
where $\ominus$ denotes substraction modulo $\mathcal{M}$. 
\item Measure all ancillas in the computational basis. Given the output $\{j_n\}_{n=2}^N$ of the measurements in step 2, we finally apply $\otimes_{n=2}^N (X_n)^{\sum_{m=2}^n j_m}$ to the qudits, where $X_n$ acts on each site as $X_n |i_n\rangle_{s_n}=|i_n\oplus 1\rangle_{s_n}$, where $\oplus$ denotes addition modulo $\mathcal{M}$.
\end{enumerate}
Here we provide an explicit expression for $U$ and $\Lambda_a$ for the case of Fibonacci fusion category. $W$ can also be constructed from $\Lambda_a$. The bulk tensor for the constructed MPDO RFPs is $M=\sum_{\alpha, ij; a,kl}\frac{\sqrt{\kronecker_\alpha}}{\mathcal{D}} R_{a,kl}^{\alpha,ij}  |\alpha,i\rangle | \alpha,j\rangle  (a,k|(a,l|$. The corresponding $U$ and $\Lambda_a$ can be obtained from the polar decomposition of $M$ as a linear map from the virtual space to the physical space:
\begin{equation}
\begin{split}
M=\sum_{\alpha,ij; a,kl}U_{a,kl}^{\alpha, ij}|\alpha,i\rangle |\alpha,j\rangle\sum_{k'}\Lambda_{a,kk'}^L(a,k' |\sum_{l'}\Lambda_{a,l'l}^R|(a,l'|,
\end{split}
\end{equation}
where $U^{\alpha,ij}_{a,kl}$ is the matrix element of $U$ and $\Lambda_a=\Lambda_a^R\Lambda_a^L$. We now write down their components accordingly. The components of $R$ is (see SM Sec.~\ref{sec:Fibonacci}~\cite{Supp}):
\begin{equation}
\begin{split}
R_{I,11}^{I,11}&=R_{I,12}^{\tau,11}=R_{\tau,11}^{I,12}=R_{I,21}^{\tau,22}=R_{\tau,22}^{I,21}=R_{I,22}^{I,22}\\
&=R_{I,22}^{\tau,33}=R_{\tau,33}^{I,22}=R_{\tau,21}^{\tau,21}=R_{\tau,23}^{\tau,31}=R_{\tau,31}^{\tau,23}=1,\\
R_{\tau,13}^{\tau,13}&=R_{\tau,32}^{\tau,32}=\zeta,\quad R_{\tau,12}^{\tau,12}=-R_{\tau,33}^{\tau,33}=\zeta^2,
\end{split}
\end{equation}
with $\zeta=\sqrt{\frac{\sqrt{5}-1}{2}}$.
Correspondingly, the components of $U$ and $\Lambda_a$ are
\begin{equation}
\begin{split}
U_{I,11}^{I,11}&=U_{I,12}^{\tau,11}=U_{\tau,11}^{I,12}=U_{I,21}^{\tau,22}=U_{\tau,22}^{I,21}=U_{\tau,12}^{\tau,12}\\
&=U_{\tau,13}^{\tau,13}=U_{\tau,21}^{\tau,21}=U_{\tau,23}^{\tau,31}=U_{\tau,31}^{\tau,23}=U_{\tau,32}^{\tau,32}=1,\\
U_{I,22}^{\tau,33}&=U_{\tau,33}^{I,22}=\zeta,\quad U_{I,22}^{I,22}=-U_{\tau,33}^{\tau,33}=\zeta^2;
\end{split}
\end{equation}
and
\begin{equation}
\Lambda_I=\frac{1}{5^{1/4}}\text{diag}\left(\zeta,\zeta^{-1}\right),\ \Lambda_\tau=\frac{1}{5^{1/4}}\text{diag}\left(\zeta,\zeta,1\right).
\end{equation}

\clearpage 

\begin{center}
\textbf{\large Supplemental Material for ``Anomalous matrix product operator symmetries and 1D mixed-state phases''}
\end{center}
\setcounter{equation}{0}
\setcounter{figure}{0}
\setcounter{table}{0}
\setcounter{page}{1}
\setcounter{section}{0}
\setcounter{defn}{0}
\setcounter{secnumdepth}{1}
\setcounter{secnumdepth}{2}
\setcounter{secnumdepth}{3}
\setcounter{lemma}{0}
\renewcommand{\thelemma}{S\arabic{lemma}}
\makeatletter
\renewcommand{\theequation}{S\arabic{equation}}
\renewcommand{\thefigure}{S\arabic{figure}}
\renewcommand{\thedefn}{S\arabic{defn}}
\renewcommand{\bibnumfmt}[1]{[S#1]}
\renewcommand{\thesection}{\Roman{section}}
\renewcommand{\thesubsection}{\Alph{subsection}}

\section{Anomaly-free condition}
\label{sec:anomaly-free}

Consider a non-anomalous MPO symmetry $\{O_a^{(L)}\}_{a\in\mathcal{B}}$ (and if necessary $\{O_a^{(L)}\}\otimes\mathbb{1}_r^{\otimes L}\}_{a\in\mathcal{B}}$) corresponding to a unitary fusion category $\mathcal{C}$, with a symmetric, translationally invariant, normal MPS $|\psi^{(L)}(A)\rangle$. According to Lemma 1 in the main text, we can always assume that $O_a^{(L)}|\psi^{(L)}(A)\rangle=n_a |\psi^{(L)}(A)\rangle$ without loss of generality, after blocking $\ell$ sites into one if needed. From the proof of Theorem 1, the Frobenius-Perron dimension of $\mathcal{L}_a$. Now let us clarify the underlying structure of a module category with a single isomorphism class of simple objects. We will not go through all details. Readers unfamiliar with unitary fusion categories are referred to Sec.~\ref{subsec:reconstruction} for an introduction.

First, we explain the idea behind how the structure of a unitary fusion category is encoded in the MPO symmetry. Let us denote the local tensor for MPO $O_a O_b$ as $T_{a\cdot b}$. It can be written explicitly as a set of matrices acting on the virtual space: $T_{a\cdot b}^{ik}=\sum_j T_a^{ij}\otimes T_b^{jk}$. Furthermore, applying the fundamental theorem~\cite{Cirac2017} to $O_a O_b=\sum_{c} N_{ab}^c O_c$, we must have
\begin{equation}
\bigoplus_{c,\mu}\mathcal{V}_{ab,\mu}^{c+} T_{a\cdot b}^{ij}\mathcal{V}_{ab,\mu}^c=\bigoplus_c\mathbb{1}_{N_{ab}^c}\otimes T_c^{ij}
\end{equation}
For the unitary fusion category case, we can find a gauge fixing the pseudo-inverse of $\mathcal{V}_{ab,\mu}$ to be its Hermitian conjugation, i.e., $\mathcal{V}_{ab,\mu}^{c+}=\mathcal{V}_{ab,\mu}^{c\dagger}$ such that $\mathcal{V}_{ab,\mu}^c$ is an isometry. One can view the normal tensors $T_{a}$ as simple objects, $T_{a\cdot b}:=T_{a}\boxtimes T_{b}$ as their monoidal products, $\mathcal{V}_{ab,\mu}^c$ as a morphism from $T_{a}\boxtimes T_{b}$ to $T_{c}$, and $\mathcal{V}_{ab,\mu}^{c\dagger}$ as the corresponding adjoint morphism from $T_{c}$ to $T_{a}\boxtimes T_{b}$, for a unitary fusion category $\mathcal{C}$. With this dictionary, all defining properties of $\mathcal{C}$ can be expressed as tensor identities (see, e.g., \cite{Bultinck2017}). 

Analogously, the local tensor condition for $O_a^{(L)}|\psi^{(L)}(A)\rangle=d_a |\psi^{(L)}(A)\rangle$ can be interpreted as encoding the structure of a module category. Let us denote the MPS tensor for $O_a^{(L)}|\psi^{(L)}(A)\rangle$ as $\tilde{A}_{a\triangleright A}$. The explicit matrix form in the virtual space is $\tilde{A}_{a\triangleright A}^{i}=\sum_{j}T_a^{ij}\otimes A^j$. From the fundamental theorem~\cite{Cirac2017}, we have
\begin{equation}
\bigoplus_{u}\mathcal{W}_{a\triangleright A,u}^{+} \tilde{A}_{a\triangleright A}^i \mathcal{W}_{a\triangleright A,u} =\mathbb{1}_{d_a}\otimes A^i.
\end{equation}
We can view the normal tensor $A$ as the only simple object in a $\mathcal{C}$-module category up to isomorphisms, $\tilde{A}_{a\triangleright A}^{i}$ as the action of $T_a$ on $A$ denoted as $T_a\odot A:= \tilde{A}_{a\triangleright A}$, and $\mathcal{W}_{a\triangleright A,u}$ as morphisms from $T_a\odot A$ to $A$. With this dictionary, all defining properties of the module category, such as the compatibility of $\odot$ with $\boxtimes$, can be expressed in the tensor language and the anomaly-free condition defined in the main text implies the existence of a module category with a single isomorphism class of simple objects. A detailed discussion about MPO symmetric MPSs and the underlying module category structure can be found in \cite{GarreRubio2023}.

\section{$C^*$-Weak Hopf algebra}
\subsection{Full definition of $C^*$-weak Hopf algebra}
\label{subsec:definition}
Here we list all axioms to define $C^*$-weak Hopf algebra which was first introduced in \cite{Bohm1996}. Let us first define the weak bialgebra. We will focus on the physically relevant case of a complex algebra.
\begin{defn}
A complex weak bialgebra is a tuple $(\mathcal{A},\mu, \eta, \Delta, \epsilon)$, where $\mathcal{A}$ is a vector space over the field of complex numbers $\mathbb{C}$, such that 

(1). $(\mathcal{A},\mu, \eta)$ forms a unital associative algebra, where $\mu: \mathcal{A}\otimes \mathcal{A}\rightarrow \mathcal{A}$ and $\eta: \mathbb{C}\rightarrow \mathcal{A}$ are linear maps satisfying the following relations:
\begin{equation}
\begin{split}
&\mu\circ(\mu\otimes \text{id})=\mu\circ(\text{id}\otimes\mu),\\
&\text{id}=\mu\circ(\eta\otimes \text{id})=\mu\circ( \text{id}\otimes \eta),
\end{split}
\end{equation}
where $\text{id}$ is the identity map from $\mathcal{A}$ to $\mathcal{A}$. For simplicity, we will denote $\mu(x\otimes y)$ as $x\cdot y$ and the unital element $\eta(1)$ as $\boldsymbol{1}$.

(2). $(\mathcal{A},\Delta, \epsilon)$ forms a unital coassociative coalgebra, where $\Delta: \mathcal{A}\rightarrow \mathcal{A}\otimes\mathcal{A}$ and $\epsilon: \mathcal{A}\rightarrow \mathbb{C}$ are linear maps satisfying the following relations:
\begin{equation}
\begin{split}
&\Delta^{(2)}:=(\Delta\otimes \text{id})\circ\Delta=(\text{id}\otimes\Delta)\circ\Delta,\\
&\text{id}=(\epsilon\otimes \text{id})\circ\Delta=(\text{id}\otimes\epsilon)\circ\Delta.
\end{split}
\end{equation}
We can further define $\Delta^{(n)}$ for all $n$ by induction.

(3). The comultiplication $\Delta$ is multiplicative, i.e.,
\begin{equation}
\Delta(x)\cdot \Delta(y)=\Delta(x\cdot y).
\end{equation}

(4). Unit axiom, i.e.,
\begin{equation}
\begin{split}
\Delta^{(2)}(\boldsymbol{1})&=(\Delta(\boldsymbol{1})\otimes \boldsymbol{1})\cdot(\boldsymbol{1}\otimes \Delta(\boldsymbol{1}))\\
&=(\boldsymbol{1}\otimes \Delta(\boldsymbol{1}))\cdot
(\Delta(\boldsymbol{1})\otimes \boldsymbol{1}).
\end{split}
\end{equation}

(5). Counit axiom, i.e., 
\begin{equation}
\epsilon(xyz)=\sum_{(y)}\epsilon(xy_{(1)})\epsilon(y_{(2)}z)=\sum_{(y)}\epsilon(xy_{(2)})\epsilon(y_{(1)}z),
\end{equation}
where we adopt Sweedler notation for the components $y_{(1)}$, $y_{(2)}$ in $\Delta(y)=\sum_{(y)}y_{(1)}\otimes y_{(2)}$. 
\end{defn}
Now we can define the complex weak Hopf algebra:
\begin{defn}
A complex weak Hopf algebra is a complex weak bialgebra $\mathcal{A}$ equipped with an additional linear map $S: \mathcal{A}\rightarrow\mathcal{A}$, called the antipode, satisfying:
\begin{equation}
\begin{split}
&\sum_{(x)}S(x_{(1)})\cdot x_{(2)}=\sum_{(\boldsymbol{1})}\boldsymbol{1}_{(1)}\epsilon(x\cdot \boldsymbol{1}_{(2)}),\\
&\sum_{(x)}x_{(1)}\cdot S(x_{(2)})=\sum_{(\boldsymbol{1})}\epsilon(\boldsymbol{1}_{(1)}\cdot x)\boldsymbol{1}_{(2)},\\
&\sum_{(x)}S(x_{(1)})\cdot x_{(2)}\cdot S(x_{(3)})=S(x),
\end{split}
\end{equation}
where we use Sweedler notation for the components $x_{(1)}, x_{(2)}, x_{(3)}$ in $\Delta^{(2)}(x)=\sum_{(x)}x_{(1)}\otimes x_{(2)}\otimes x_{(3)}$.
\end{defn}

We will focus on the finite-dimensional case and a finite-dimensional complex $C^*$-weak Hopf algebra is defined as:
\begin{defn}
A finite-dimensional complex $C^*$-weak Hopf algebra is a finite-dimensional weak complex bialgebra equipped with an anti-linear map $*: \mathcal{A}\rightarrow\mathcal{A}$ such that it is an involution ($x^{**}=x$), antihomomorphism ($(x\cdot y)^*=y^*\cdot x^* $), cohomomorphism ($\Delta\circ*=(*\otimes*)\circ\Delta$), and adopt a faithful *-representation.
\end{defn}
Here a *-representation is representation $\varphi$ such that $\varphi(x^*)=\varphi^{\dagger}(x)$ on a complex Hilbert space and it is faithful when the map from the algebra to the complex Hilbert space is injective. A finite-dimensional complex $C^*$-weak Hopf algebra $\mathcal{A}$ has its dual algebra $\mathcal{A}^*$ defined on the dual linear space, i.e., linear functions $f(x)$ on $\mathcal{A}$. $\mathcal{A}^*$ is also a $C^*$-weak Hopf algebra equipped with multiplication $\mu$, unit $\tilde{\eta}$, comultiplication $\Delta$, counit $\epsilon$, antipode $S$, and a *-operation defined as follows:
\begin{equation}
\begin{split}
&\mu\circ (f_1\otimes f_2):=(f_1\otimes f_2)\circ \Delta,\\
&\tilde{\eta}(z):=z\epsilon,\quad z\in\mathbb{C},\\
&\Delta(f):=f\circ \mu,\\
&\epsilon(f):=f(\boldsymbol{1}),\\
&S(f):=f\circ S,\\
&f^*:=K\circ f \circ *\circ S,
\end{split}
\label{eq: dual_operations}
\end{equation}
where $K$ denotes the complex conjugation operator on $\mathbb{C}$. We use the same notation for multiplication, comultiplication, counit, antipode, and the *-operation in both $\mathcal{A}$ and $\mathcal{A}^*$, as the argument always makes the domain clear. For the unit maps, we adopt different notations since both act on $\mathbb{C}$.

\subsection{Unitary fusion category and reconstruction of $C^*$-weak Hopf algebra}
\label{subsec:reconstruction}
One can construct a weak Hopf algebra $\mathcal{A}$ from an invertible $(\mathcal{C}_1, \mathcal{C}_2)$ bimodule category $\mathcal{M}$ for two fusion categories $\mathcal{C}_1$ and $\mathcal{C}_2$ such that $\text{Rep}(\mathcal{A}^*)\cong\mathcal{C}_1$ and $\text{Rep}(\mathcal{A})\cong\mathcal{C}_2$~\cite{Ostrik2003,Etingof2015}. A natural choice that always exists is to view any fusion category $\mathcal{C}$ as a $(\mathcal{C},\mathcal{C})$ bimodule category and one can recover a weak Hopf algebra $\mathcal{A}$ such that $\text{Rep}(\mathcal{A}^*)\cong \text{Rep}(\mathcal{A})\cong \mathcal{C}$. In this case, we can use a diagramatic approach~\cite{Kitaev2012} to explicitly construct a $C^*$-weak Hopf algebra $\mathcal{A}$ such that $\text{Rep}(\mathcal{A}^*)\cong \mathcal{C}$.

We start from the basics of category theory and introduce the diagramatics for the fusion category\footnote{The reader is referred to \cite{Kitaev2006} for a more complete introduction, which covers most part of the discussions on unitary fusion category.}. A category consists of objects and morphisms between them that are associative and contain an identity morphism for each object. To define unitary fusion category, let us first introduce a unitary category as equivalent to the category of formal sums $X=\bigoplus_{a\in \mathcal{B}}X_a \cdot a$, where $\mathcal{B}$ is a set of labels and $\{X_a\}$ consists of finite-dimensional Hilbert spaces of which only finitely many are nonzero. A morphism $f\in \text{Hom}_{\mathcal{C}}(X,Y)$ is a collection of linear maps $f_a: X_a \rightarrow Y_a$, which forms a linear space. The adjoint morphism $f^{\dagger}$ is defined as the collection of $f_a^{\dagger}: Y_a \rightarrow X_a$. The simple object $[a]$ is defined as the object such that $[a]_a=\mathbb{C}$, $[a]_b=0$ for $b\neq a$, and $X_a \cong \text{Hom}_{\mathcal{C}}([a],X)$ as a Hilbert space. The identity morphism for $X$, i.e., $\text{id}_X$ is the collection of the identity linear maps on $X_i$.

Let us consider the case where $\mathcal{B}$ is a finite set, and a unitary fusion category $\mathcal{C}$ can be viewed as a unitary category with additional structures. For a unitary fusion category, there exists a monoidal product of two objects $X$ and $Y$: $X\boxtimes Y$, such that $(X\boxtimes Y)_c=\bigoplus_{a,b}V_c^{ab}\otimes X_a \otimes Y_b$, where $V_c^{ab}\cong \text{Hom}_{\mathcal{C}}([c],[a]\boxtimes[b])$ as a Hilbert space, and is called the splitting space. This monoidal product is a functor, meaning for any morphisms $f$ and $g$, we can also define $f\boxtimes g$ such that: 
\begin{equation}
\begin{split}
(f_2 \circ f_1)\boxtimes(g_2\circ g_1)&=(f_2\boxtimes g_2)\circ(f_1\boxtimes g_1), \\
\text{id}_{X}\boxtimes \text{id}_{Y} &=\text{id}_{X\boxtimes Y}.
\end{split}
\end{equation}
Here the definition $f\boxtimes g$ can be explicitly written as $(f\boxtimes g)_c=\sum_{a,b}\text{id}_{V_c^{ab}}\otimes f_a\otimes g_b$. We can also define the dual of $V_c^{ab}$: $V_{ab}^c\cong \text{Hom}_{\mathcal{C}}([a]\boxtimes[b],[c])$. The dimension of $V_{c}^{ab}$ and $V_{ab}^c$ is denoted by $N_{ab}^c$ and the quantum dimension $d_a$ of $[a]$ coincides with the largest eigenvalues of the integer matrix $(N_a)_{cb}:=N_{ab}^c$. We can write bases of $V_c^{ab}$ and $V_{ab}^c$ diagrammatically as
\begin{equation}
\begin{array}{c}
\begin{tikzpicture}[scale=.5, baseline={([yshift=0ex]current bounding box.center)}, thick]
\splitspace{(0,0)}{$a$}{$b$}{$c$}{$\mu$}
\end{tikzpicture}
\end{array}
\equiv|\mu\rangle\in V_c^{ab},\quad
\begin{array}{c}
\begin{tikzpicture}[scale=.5, baseline={([yshift=0ex]current bounding box.center)}, thick]
\fusionspace{(0,0)}{$a$}{$b$}{$c$}{$\nu^*$}
\end{tikzpicture}
\end{array}
\equiv\langle\nu|\in V^c_{ab}.
\end{equation}
We can also compose these morphisms and this amounts to stacking diagrams from bottom to top, e.g.,
\begin{equation}
\begin{array}{c}
\begin{tikzpicture}[scale=.5, baseline={([yshift=0ex]current bounding box.center)}, thick]
\draw (0,1) -- (0,2);
\draw (0,-1) to[out=30,in=-30] (0,1);
\draw (0,-1) to[out=150,in=-150] (0,1);
\draw (0,-1) -- (0,-2);
\draw (0.5,1.5) node {\small $\nu^*$};
\draw (0.4,-1.4) node {\small $\mu$};
\draw (-1,0) node {\small $a$};
\draw (1,0) node {\small $b$};
\draw (0,2.4) node {\small $c$};
\draw (0,-2.4) node {\small $c$};
\end{tikzpicture}
\end{array}=
\langle\nu|\mu\rangle 
\begin{array}{c}
\begin{tikzpicture}[scale=.5, baseline={([yshift=0ex]current bounding box.center)}, thick]
\draw (0,-2) -- (0,2);
\draw (0,2.4) node {\small $c$};
\draw (0,-2.4) node {\small $c$};
\end{tikzpicture}
\end{array},\quad \text{with}
\begin{array}{c}
\begin{tikzpicture}[scale=.5, baseline={([yshift=0ex]current bounding box.center)}, thick]
\draw (0,-2) -- (0,2);
\draw (0,2.4) node {\small $c$};
\draw (0,-2.4) node {\small $c$};
\end{tikzpicture}
\end{array}
\equiv \text{id}_{[c]}.
\end{equation}
It is convenient to normalize the basis vectors such that:
\begin{equation}
    \langle\nu|\mu\rangle=\delta_{\mu\nu}\sqrt{\frac{d_a d_b}{d_c}}.
\end{equation}
There is a completeness relation which expands $\text{id}_{[a]\boxtimes [b]}=\text{id}_{[a]}\boxtimes \text{id}_{[b]}$ with the composition of the above basis:
\begin{equation}
\text{id}_{[a]}\boxtimes \text{id}_{[b]}\equiv
\begin{array}{c}
\begin{tikzpicture}[scale=.5, baseline={([yshift=0ex]current bounding box.center)}, thick]
\draw (-1,-2) -- (-1,2);
\draw (-1,2.4) node {\small $a$};
\draw (-1,-2.4) node {\small $a$};
\draw (1,-2) -- (1,2);
\draw (1,2.4) node {\small $b$};
\draw (1,-2.4) node {\small $b$};
\end{tikzpicture}
\end{array}
=\sum_{c\mu}\sqrt{\frac{d_c}{d_a d_b}}\begin{array}{c}
\begin{tikzpicture}[scale=.5, baseline={([yshift=0ex]current bounding box.center)}, thick]
\splitspace{(0,1)}{$a$}{$b$}{$c$}{$\mu^*$}
\fusionspace{(0,-1)}{$a$}{$b$}{$c$}{$\mu$};
\end{tikzpicture}
\end{array}.
\end{equation}
We require the existence of a special object $[I]$ with $I\in \mathcal{B}$ called the unit object and the existence of isomorphisms: $F_{X,Y,Z}:=(X\boxtimes Y)\boxtimes Z\rightarrow X\boxtimes(Y\boxtimes Z)$ with $(F_{X,Y,Z})_{d}=\sum_{a,b,c}F_d^{abc}\otimes \text{id}_{X_a}\otimes\text{id}_{Y_b}\otimes\text{id}_{Z_c}$, $\alpha_X: X\rightarrow X\boxtimes [I]$ and $\beta_X: X\rightarrow [I]\boxtimes X$, which satisfy the following pentagon equation
\begin{equation}
\begin{split}
&\left(\text{id}_X\boxtimes F_{Y,Z,W}\right)\circ F_{X,Y\boxtimes Z, W}\circ\left(F_{X,Y,Z}\boxtimes \text{id}_W\right)\\
&= F_{X,Y,Z\boxtimes W}\circ F_{X\boxtimes Y,Z,W},
\end{split}
\end{equation}
and triangle equation:
\begin{equation}
\text{id}_X\boxtimes \beta_Y= F_{X,[I],Y}\circ (\alpha_X\boxtimes \text{id}_Y).
\end{equation}
 Here we have used $F_{d}^{abc}$ to denote a canonical unitary isomorphism: $\left(([a]\boxtimes [b])\boxtimes [c]\right)_d\rightarrow \left([a]\boxtimes ([b]\boxtimes [c])\right)_d$.
These isomorphisms $(F,\alpha,\beta)$ are natural, i.e., given morphisms $f: X\rightarrow X'$, $g: Y\rightarrow Y'$, and $h: Z\rightarrow Z'$, we have
\begin{equation}
\begin{split}
&F_{X',Y',Z'}\circ\left((f\boxtimes g)\boxtimes h \right)=\left( f\boxtimes (g\boxtimes h)\right)\circ F_{X,Y,Z},\\
&\alpha_{X'}\circ f=(f\boxtimes \text{id}_{[I]})\circ \alpha_X,\quad \beta_{X'} \circ f=(\text{id}_{[I]}\boxtimes f)\circ \beta_X.
\end{split}
\end{equation}
Notice that 
\begin{equation}
\begin{split}
&\left(([a]\boxtimes [b])\boxtimes [c]\right)_d=\bigoplus_e V^{ab}_e\otimes V_{d}^{ec},\\
&\left([a]\boxtimes ([b]\boxtimes [c])\right)_d=\bigoplus_f V^{af}_d\otimes V_{f}^{bc},
\end{split}
\end{equation} 
and the isomorphism between them can be written in terms of bases for both Hilbert spaces, which defines the $F$-symbols:
\begin{equation}
\begin{array}{c}
\begin{tikzpicture}[scale=.5, baseline={([yshift=0ex]current bounding box.center)}, thick]
\draw (0,0) -- (0,1);
\draw (0,1) -- (-2,3);
\draw (-1,2) -- (0,3);
\draw (0,1) -- (2,3);
\draw (0,-0.5) node{$d$};
\draw (-1,2.6) node{$\mu$};
\draw (-0.9,1.4) node{$e$};
\draw (0,1.6) node{$\nu$};
\draw (-2,3.5) node{$a$};
\draw (0,3.5) node{$b$};
\draw (2,3.5) node{$c$};
\end{tikzpicture}
\end{array}
=\sum_{f\rho\lambda}[F_{d}^{abc}]_{f\rho\lambda}^{e\mu\nu}
\begin{array}{c}
\begin{tikzpicture}[scale=.5, baseline={([yshift=0ex]current bounding box.center)}, thick]
\draw (0,0) -- (0,1);
\draw (0,1) -- (2,3);
\draw (1,2) -- (0,3);
\draw (0,1) -- (-2,3);
\draw (0,-0.5) node{\small $d$};
\draw (1,2.6) node{\small $\rho$};
\draw (0.9,1.4) node{\small $f$};
\draw (0,1.6) node{\small $\lambda$};
\draw (-2,3.5) node{\small $a$};
\draw (0,3.5) node{\small $b$};
\draw (2,3.5) node{\small $c$};
\end{tikzpicture}
\end{array}.
\end{equation}
where $[F_{d}^{abc}]_{f\rho\lambda}^{e\mu\nu}:=(\langle \lambda|\otimes\langle \rho|) |F_d^{abc}(|\mu\rangle\otimes |\nu\rangle )$, with $|\mu\rangle\otimes|\nu\rangle \in V_e^{ab}\otimes V_d^{ec}$ and $|\lambda\rangle\otimes|\rho\rangle\in V_d^{af}\otimes V_f^{bc}$. Furthermore, the natural isomorphisms $\alpha$ and $\beta$ indicate that $V_{a}^{aI}$ and $V_{a}^{Ia}$ are canonically isomorphic to $\mathbb{C}$ with fixed basis vectors $|\alpha_a\rangle$ and $|\beta_a\rangle$, respectively. The triangle equation requires that $F_{c}^{aIb}\left(|\alpha_a\rangle\otimes |\mu\rangle\right)=|\mu\rangle\otimes |\beta_b\rangle$ for any $|\mu\rangle\in V_{c}^{ab}$. 

One additional structure for the unitary fusion category is that every object admits a two-sided dual, namely, for every object $X$, there exists $\overline{X}$ together with evaluation and coevaluation morphisms:  $\text{ev}_X: \overline{X}\boxtimes X\rightarrow I$, and $\text{coev}_X: I\rightarrow X\boxtimes \overline{X}$, such that the zig-zag identity holds
\begin{equation}
\begin{split}
\text{id}_X&=(\text{id}_X\boxtimes \text{ev}_X)\circ (\text{coev}_X\boxtimes \text{id}_X),\\
\text{id}_{\overline{X}}&=(\text{ev}_X\boxtimes\text{id}_{\overline{X}})\circ (\text{id}_{\overline{X}}\boxtimes \text{coev}_X).
\end{split}
\end{equation} 
The above property means that $\overline{X}$ is the left dual of $X$. For unitary fusion category, $\overline{X}$ is also a right dual of $X$ equipped with morphisms $\text{ev}_X^{\dagger}: I\rightarrow \overline{X}\boxtimes X$, and $\text{coev}_X^\dagger:=X\boxtimes\overline{X}\rightarrow I$ such that a similar zig-zag identity holds. We can use the following diagram for simple object $[a]$ and its dual $[\bar{a}]$ to represent morphisms associated the duality property:
\begin{equation}
\begin{split}
\text{coev}_{[a]}&=
\begin{array}{c}
\begin{tikzpicture}[scale=.5, baseline={([yshift=0ex]current bounding box.center)}, thick]
\draw (0,0) -- (1,1);
\draw (0,0) -- (-1,1);
\draw[dotted] (0,0) -- (0,-1);
\draw (-1.4,1) node {\small $a$};      
\draw (1.4,1) node {\small $\bar{a}$};      
\draw (0.4,-1) node {\small $I$};
\end{tikzpicture}
\end{array}
=\begin{array}{c}
\begin{tikzpicture}[scale=.5, baseline={([yshift=0ex]current bounding box.center)}, thick]
\draw (0,0)  arc (0:-180:1.2);
\draw[fill=white] (-1.3,-1)--(-1.3,-1.4)--(-0.9,-1.2)--(-1.3,-1);
\draw (-2.4,0.4) node {\small $a$};      
\draw (0,0.4) node {\small $\bar{a}$};      
\end{tikzpicture}
\end{array}\\
&=\kappa_a
\begin{array}{c}
\begin{tikzpicture}[scale=.5, baseline={([yshift=0ex]current bounding box.center)}, thick]
\draw (0,0)  arc (0:-180:1.2);
\draw[fill=white] (-0.9,-1)--(-0.9,-1.4)--(-1.3,-1.2)--(-0.9,-1);
\draw (-2.4,0.4) node {\small $a$};      
\draw (0,0.4) node {\small $\bar{a}$};    
\end{tikzpicture}
=\kappa_a \text{ev}_{[\bar{a}]}^{\dagger},
\end{array},
\end{split}
\end{equation}
and the Hermitian conjugations of above diagrams:
\begin{equation}
\begin{array}{c}
\begin{tikzpicture}[scale=.5, baseline={([yshift=0ex]current bounding box.center)}, thick]
\draw (0,0) -- (1,-1);
\draw (0,0) -- (-1,-1);
\draw[dotted] (0,0) -- (0,1);
\draw (-1.4,-1) node {\small $a$};      
\draw (1.4,-1) node {\small $\bar{a}$};      
\draw (0.4,1) node {\small $I$};
\end{tikzpicture}
\end{array}
=\begin{array}{c}
\begin{tikzpicture}[scale=.5, baseline={([yshift=0ex]current bounding box.center)}, thick]
\draw (0,0)  arc (0:180:1.2);
\draw[fill=white] (-1.3,1)--(-1.3,1.4)--(-0.9,1.2)--(-1.3,1);
\draw (-2.4,-0.4) node {\small $a$};      
\draw (0,-0.4) node {\small $\bar{a}$};   
\end{tikzpicture}
\end{array}
=\kappa_a^*
\begin{array}{c}
\begin{tikzpicture}[scale=.5, baseline={([yshift=0ex]current bounding box.center)}, thick]
\draw (0,0)  arc (0:180:1.2);
\draw[fill=white] (-0.9,1)--(-0.9,1.4)--(-1.3,1.2)--(-0.9,1);
\draw (-2.4,-0.4) node {\small $a$};      
\draw (0,-0.4) node {\small $\bar{a}$};   
\end{tikzpicture}
\end{array},
\end{equation}
where $\kappa_a$ is the Frobenius-Schur indicator (a phase factor) relating the two morphisms $\text{coev}_{[a]}$ and $\text{ev}^{\dagger}_{[\bar{a}]}: I\rightarrow [a]\boxtimes[\bar{a}]$.  The zig-zag identities can be written in terms of diagrams:
\begin{equation}
\begin{array}{c}
\begin{tikzpicture}[scale=.5, baseline={([yshift=0ex]current bounding box.center)}, thick]
\draw (2.4,0)  arc (0:180:1.2);
\draw[fill=white] (1.5,1)--(1.5,1.4)--(1.1,1.2)--(1.5,1);
\draw (-2,0) node {\small $a$};
\draw (0.4,0) node {\small $\bar{a}$};
\draw (0,0)  arc (0:-180:1.2);
\draw[fill=white] (-1.3,-1)--(-1.3,-1.4)--(-0.9,-1.2)--(-1.3,-1);      
\draw (2.8,0) node {\small $a$};   
\end{tikzpicture}
\end{array}
=\begin{array}{c}
\begin{tikzpicture}[scale=.5, baseline={([yshift=0ex]current bounding box.center)}, thick]
\draw (0,0) -- (0,2);     
\draw (0.4,1) node {\small $a$};   
\end{tikzpicture}
\end{array}
=\begin{array}{c}
\begin{tikzpicture}[scale=.5, baseline={([yshift=0ex]current bounding box.center)}, thick]
\draw (2.4,0)  arc (0:-180:1.2);
\draw[fill=white] (1.5,-1)--(1.5,-1.4)--(1.1,-1.2)--(1.5,-1);
\draw (-2,0) node {\small $a$};
\draw (0.4,0) node {\small $\bar{a}$};
\draw (0,0)  arc (0:180:1.2);
\draw[fill=white] (-1.3,1)--(-1.3,1.4)--(-0.9,1.2)--(-1.3,1);      
\draw (2.8,0) node {\small $a$};   
\end{tikzpicture}
\end{array}.
\end{equation}
In later context, as a convention, when the ``cups'' and ``caps'' diagrams are connected with opposite flags, these flags will be left implicit. 

Now with the above data from a unitary fusion category $\mathcal{C}$, we are able to construct a corresponding $C^*$-weak Hopf algebra. This amounts to the diagramatic construction introduced in \cite{Kitaev2012} for studying gapped boundaries when choosing the boundary module category to be $\mathcal{C}$ itself. The basis of the algebra contains:
\begin{equation}
\begin{tikzpicture}[scale=.5, baseline={([yshift=-0.8ex]current bounding box.center)}, thick]
\draw (0,0)--(0,1.5);
\draw (0,3)--(0,4.5);
\draw (0,0.75) to[out=150,in=-150] (0,3.75);
\draw (-1,2.25) node {\small $a$};   
\draw (0.5,0) node {\small $c_1$}; 
\draw (0.5,1.5) node {\small $c_2$};   
\draw (0.5,0.75) node {\small $\mu$};  
\draw (0.5,3) node {\small $d_2$};   
\draw (0.5,4.5) node {\small $d_1$};   
\draw (0.5,3.75) node {\small $\nu^*$};  
\end{tikzpicture},
\quad \{a,c_1,c_2,d_1,d_2\}\in \mathcal{B},
\end{equation}
We can define the required structures, i.e., multiplication, comultiplication, unit, counit, antipode, and *-operation, for the $C^*$-weak Hopf algebra as follows:
\begin{equation}
\left[\begin{tikzpicture}[scale=.5, baseline={([yshift=-0.8ex]current bounding box.center)}, thick]
\draw (0,0)--(0,1.5);
\draw (0,3)--(0,4.5);
\draw (0,0.75) to[out=150,in=-150] (0,3.75);
\draw (-1,2.25) node {\small $a$};   
\draw (0.5,0) node {\small $c_1$}; 
\draw (0.5,1.5) node {\small $c_2$};   
\draw (0.5,0.75) node {\small $\mu$};  
\draw (0.5,3) node {\small $d_2$};   
\draw (0.5,4.5) node {\small $d_1$};   
\draw (0.5,3.75) node {\small $\nu^*$};  
\end{tikzpicture}
\right]
\boldsymbol{\cdot}
\left[
\begin{tikzpicture}[scale=.5, baseline={([yshift=-0.8ex]current bounding box.center)}, thick]
\draw (0,0)--(0,1.5);
\draw (0,3)--(0,4.5);
\draw (0,0.75) to[out=150,in=-150] (0,3.75);
\draw (-1,2.25) node {\small $b$};   
\draw (0.5,0) node {\small $c_2'$}; 
\draw (0.5,1.5) node {\small $c_3$};  
\draw (0.5,0.75) node {\small $\rho$};  
\draw (0.5,3) node {\small $d_3$};   
\draw (0.5,4.5) node {\small $d_2'$};   
\draw (0.5,3.75) node {\small $\lambda^*$};  
\end{tikzpicture}
\right]
=\delta_{d_2d_2'}\delta_{c_2c_2'}
\left[
\begin{tikzpicture}[scale=.5, baseline={([yshift=-0.8ex]current bounding box.center)}, thick]
\draw (0,0)--(0,1.5);
\draw (0,3)--(0,4.5);
\draw (0,0.75) to[out=150,in=-150] (0,3.75);
\draw (-1,2.25) node {\small $b$};   
\draw (0.5,0) node {\small $c_2$}; 
\draw (0.5,1.5) node {\small $c_3$};   
\draw (0.5,0.75) node {\small $\rho$};    
\draw (0.5,3) node {\small $d_3$};   
\draw (0.5,4.5) node {\small $d_2$};   
\draw (0.5,3.75) node {\small $\lambda^*$};  
\draw (0,-1.5)--(0,0);
\draw (0,4.5)--(0,6);
\draw (0,-0.75) to[out=150,in=-150] (0,5.25);
\draw (-1.8,2.25) node {\small $a$};   
\draw (0.5,-1.5) node {\small $c_1$}; 
\draw (0.5,-0.75) node {\small $\mu$};    
\draw (0.5,6) node {\small $d_1$};   
\draw (0.5,5.25) node {\small $\nu^*$};    
\end{tikzpicture}
\right],
\label{eq:mul}
\end{equation}
\begin{equation}
{\bf 1}=\sum_{c,d}
\left[\ \begin{tikzpicture}[scale=.5, baseline={([yshift=-0.8ex]current bounding box.center)}, thick]
\draw (0,0)--(0,1.5);
\draw (0,3)--(0,4.5);
\draw (0.5,0.75) node {\small $d$}; 
\draw (0.5,3.75) node {\small $c$};   
\end{tikzpicture}
\right],
\label{eq:unit}
\end{equation}
\begin{equation}
\Delta \left[
\begin{tikzpicture}[scale=.5, baseline={([yshift=-0.8ex]current bounding box.center)}, thick]
\draw (0,0)--(0,1.5);
\draw (0,3)--(0,4.5);
\draw (0,0.75) to[out=150,in=-150] (0,3.75);
\draw (-1,2.25) node {\small $a$};   
\draw (0.5,0) node {\small $c_1$}; 
\draw (0.5,1.5) node {\small $c_2$};   
\draw (0.5,0.75) node {\small $\mu$};  
\draw (0.5,3) node {\small $d_2$};   
\draw (0.5,4.5) node {\small $d_1$};   
\draw (0.5,3.75) node {\small $\nu^*$};  
\end{tikzpicture}
\right]=\sum_{e_1,e_2, \rho}\sqrt{\frac{d_{e_1}}{d_a d_{e_2}}}
\left[
\begin{tikzpicture}[scale=.5, baseline={([yshift=-0.8ex]current bounding box.center)}, thick]
\draw (0,0)--(0,1.5);
\draw (0,3)--(0,4.5);
\draw (0,0.75) to[out=150,in=-150] (0,3.75);
\draw (-1,2.25) node {\small $a$};   
\draw (0.5,0) node {\small $e_1$}; 
\draw (0.5,1.5) node {\small $e_2$};   
\draw (0.5,0.75) node {\small $\rho$};  
\draw (0.5,3) node {\small $d_2$};   
\draw (0.5,4.5) node {\small $d_1$};   
\draw (0.5,3.75) node {\small $\nu^*$};  
\end{tikzpicture}
\right]\otimes
\left[
\begin{tikzpicture}[scale=.5, baseline={([yshift=-0.8ex]current bounding box.center)}, thick]
\draw (0,0)--(0,1.5);
\draw (0,3)--(0,4.5);
\draw (0,0.75) to[out=150,in=-150] (0,3.75);
\draw (-1,2.25) node {\small $a$};   
\draw (0.5,0) node {\small $c_1$}; 
\draw (0.5,1.5) node {\small $c_2$};   
\draw (0.5,0.75) node {\small $\mu$};  
\draw (0.5,3) node {\small $e_2$};   
\draw (0.5,4.5) node {\small $e_1$};   
\draw (0.5,3.75) node {\small $\rho^*$};  
\end{tikzpicture}
\right],
\label{eq:comul}
\end{equation}
\begin{equation}
\epsilon \left[
\begin{tikzpicture}[scale=.5, baseline={([yshift=-0.8ex]current bounding box.center)}, thick]
\draw (0,0)--(0,1.5);
\draw (0,3)--(0,4.5);
\draw (0,0.75) to[out=150,in=-150] (0,3.75);
\draw (-1,2.25) node {\small $a$};   
\draw (0.5,0) node {\small $c_1$}; 
\draw (0.5,1.5) node {\small $c_2$};   
\draw (0.5,0.75) node {\small $\mu$};  
\draw (0.5,3) node {\small $d_2$};   
\draw (0.5,4.5) node {\small $d_1$};   
\draw (0.5,3.75) node {\small $\nu^*$};  
\end{tikzpicture}
\right]=\delta_{c_1 d_1}\delta_{c_2 d_2}\sqrt{\frac{d_a d_{c_2}}{d_{c_1}}}\delta_{\mu\nu}
\label{eq:counit}
\end{equation}
\begin{equation}
S \left[\begin{tikzpicture}[scale=.5, baseline={([yshift=-0.8ex]current bounding box.center)}, thick]
\draw (0,0)--(0,1.5);
\draw (0,3)--(0,4.5);
\draw (0,0.75) to[out=150,in=-150] (0,3.75);
\draw (-1,2.25) node {\small $a$};   
\draw (0.5,0) node {\small $c_1$}; 
\draw (0.5,1.5) node {\small $c_2$};   
\draw (0.5,3) node {\small $d_2$};   
\draw (0.5,4.5) node {\small $d_1$};   
\draw (0.5,0.75) node {\small $\mu$};  
\draw (0.5,3.75) node {\small $\nu^*$};  
\end{tikzpicture}
\right]=\frac{d_{d_2}}{d_{d_1}}\left[
\begin{tikzpicture}[scale=.5, baseline={([yshift=-0.8ex]current bounding box.center)}, thick]
\draw (0,0)--(0,1.5);
\draw (0,3)--(0,4.5);
\draw (-1,0.75)--(-1,3.75);
\draw (0,0.75) arc (0:-180:0.5);
\draw (0,3.75) arc (0:180:0.5);
\draw (-1.3,2.25) node {\small $\bar{a}$};   
\draw (0.5,0) node {\small $d_2$}; 
\draw (0.5,1.5) node {\small $d_1$};   
\draw (0.5,3) node {\small $c_1$};   
\draw (0.5,4.5) node {\small $c_2$};   
\draw (0.5,0.75) node {\small $\nu^*$};  
\draw (0.5,3.75) node {\small $\mu$};  
\end{tikzpicture}
\right]
\label{eq:antipode}
\end{equation}
\begin{equation}
\left[\begin{tikzpicture}[scale=.5, baseline={([yshift=-0.8ex]current bounding box.center)}, thick]
\draw (0,0)--(0,1.5);
\draw (0,3)--(0,4.5);
\draw (0,0.75) to[out=150,in=-150] (0,3.75);
\draw (-1,2.25) node {\small $a$};   
\draw (0.5,0) node {\small $c_1$}; 
\draw (0.5,1.5) node {\small $c_2$};   
\draw (0.5,3) node {\small $d_2$};   
\draw (0.5,4.5) node {\small $d_1$};   
\draw (0.5,0.75) node {\small $\mu$};  
\draw (0.5,3.75) node {\small $\nu^*$};  
\end{tikzpicture}
\right]^*=\frac{d_{c_2}}{d_{c_1}}\left[\begin{tikzpicture}[scale=.5, baseline={([yshift=-0.8ex]current bounding box.center)}, thick]
\draw (0,0)--(0,1.5);
\draw (0,3)--(0,4.5);
\draw (-1,0.75)--(-1,3.75);
\draw (0,0.75) arc (0:-180:0.5);
\draw (0,3.75) arc (0:180:0.5);
\draw (-1.3,2.25) node {\small $\bar{a}$};   
\draw (0.5,0) node {\small $c_2$}; 
\draw (0.5,1.5) node {\small $c_1$};   
\draw (0.5,3) node {\small $d_1$};   
\draw (0.5,4.5) node {\small $d_2$};   
\draw (0.5,0.75) node {\small $\mu^*$};  
\draw (0.5,3.75) node {\small $\nu$};  
\end{tikzpicture}\right]
\label{eq:star}
\end{equation}
The diagrams on the right hand sides of these equations can be transformed back to the basis with $F$-symbols and the inclusion of additional lines for the unit object $I$ when necessary. The verification of the defining properties of $C^*$-weak Hopf algebra is similar to \cite{Jia2024}, with appropriate modifications for the case without planar isotopy invariance. According to \cite{Kitaev2012}, we have $\text{Rep}(\mathcal{A})\cong \text{Fun}_{\mathcal{C}}(\mathcal{C},\mathcal{C})^{\text{rev}}\cong\mathcal{C}$, where $\cong$ denotes monoidal equivalence and $\text{Fun}_{\mathcal{C}}(\mathcal{C},\mathcal{C})^{\text{rev}}$ denotes the category of (left) $\mathcal{C}$-module endofunctors on $\mathcal{C}$, with monoidal product defined via functor composition in reverse order. $\text{Rep}(\mathcal{A}^*)$ can be obtained by considering the weak Hopf algebra $\widetilde{\mathcal{A}}$ defined using basis:
\begin{equation}
\begin{tikzpicture}[scale=.5, baseline={([yshift=-0.8ex]current bounding box.center)}, thick]
\draw (0,0)--(0,1.5);
\draw (0,3)--(0,4.5);
\draw (0,0.75) to[out=30,in=-30] (0,3.75);
\draw (1,2.25) node {\small $a$};   
\draw (-0.5,0) node {\small $c_1$}; 
\draw (-0.5,1.5) node {\small $c_2$};   
\draw (-0.5,0.75) node {\small $\mu$};  
\draw (-0.5,3) node {\small $d_2$};   
\draw (-0.5,4.5) node {\small $d_1$};   
\draw (-0.5,3.75) node {\small $\nu^*$};  
\end{tikzpicture},
\quad \{a,c_1,c_2,d_1,d_2\}\in \mathcal{B},
\end{equation}
with rules for multiplication, comultiplication, unit, counit, antipode, and *-operation defined similar to Eqs.~(\ref{eq:mul}-\ref{eq:star}) with all diagrams flipped horizontally. The representation category $\text{Rep}(\widetilde{\mathcal{A}})$, is monoidally equivalent to $\mathcal{C}^{\text{rev}}$, the fusion category obtained from $\mathcal{C}$ by reversing the monoidal product. A (skew) pairing introduced in \cite{Bai2025} between $\mathcal{A}$ and $\widetilde{\mathcal{A}}$ then leads to $\text{Rep}(\mathcal{A^*})\cong\mathcal{C}$.

\section{Additional details on MPDO RFPs constructed from $C^*$-weak Hopf algebra}

\subsection{Dual representations and $O_{\bar{a}}^{(L)}=O_a^{(L)\dagger}$}
\label{subsec:dual_rep}
Let us consider the representation of $\mathcal{A}^*$. The dual representation $\Psi_{\bar{a}}$ for the *-irrep $\Psi_a$ satisfies
\begin{equation}
\Psi_{\bar{a}}\cong \Psi_a^T\circ S,
\end{equation}
where $\Psi_a^T$ is defined through the transpose of the representation matrix, i.e., $\Psi_a^T(\delta_x)=[\Psi_a(\delta_x)]^T$ for all $\delta_x\in\mathcal{A}^*$. As can be verified from properties of antipode, $\Psi_a^T\circ S$ is also an irrep and the characters of irreps $a$ and $\bar{a}$ are related by $\chi_{\bar{a}}(\delta_x)=\chi_a(S(\delta_x))$.
Notice that 
\begin{equation}
\begin{split}
\sum_{x\in B_{\mathcal{A}}} x^* \delta^*_x (y)&=\sum_{x\in B_{\mathcal{A}}} x^*\overline{\delta_x[S(y)^*]}=S(y)\\
&=\sum_{x\in B_{\mathcal{A}}} x\delta_x[S(y)]=\sum_{x\in B_{\mathcal{A}}} x [S(\delta_x)](y),
\end{split}
\end{equation}
holds for all $y$ where we have used properties of the antipode and *-operation of $\mathcal{A}^*$ (see Eq.~\eqref{eq: dual_operations}). Hence we have 
\begin{equation}
\sum_{x\in B_{\mathcal{A}}} x^*\otimes \delta^*_x=\sum_{x\in B_{\mathcal{A}}} x\otimes S(\delta_x).
\label{eq: antipode_basis}
\end{equation}
Applying $\Phi^{\boxtimes L}\otimes \chi_a$ to both sides of Eq.~\eqref{eq: antipode_basis} and using the fact that $\Phi$ is a *-representation, we obtain
\begin{equation}
\begin{split}
O_{a}^{(L)\dagger}&=\sum_{x\in B_{\mathcal{A}}}\Phi^{\boxtimes L}(x^*)\chi_{a}(\delta_{x^*})\\
&=\sum_{x\in B_{\mathcal{A}}}\Phi^{\boxtimes L}(x)\chi_{a}(S(\delta_{x}))\\
&=\sum_{x\in B_{\mathcal{A}}}\Phi^{\boxtimes L}(x)\chi_{\bar{a}}(\delta_{x})= O_{\bar{a}}^{(L)}.
\end{split}
\end{equation}

\subsection{Proof of zero correlation length and local indistinguishability}
\label{subsec:trace-one-site}
Consider an MPDO fixed point generated by tensor $M=\sum_{x\in B_{\mathcal{A}}}[\Omega\Phi(x)]\otimes\Psi(\delta_x)$ and boundary matrix $\mathcal{X}=\sum_a\mathcal{X}_aP_a$ as defined in the main text. The MPDO after tracing out one site is $\rho^{(L-1)}(M, \mathcal{X}E)$ with $E=\sum_{x\in B_{\mathcal{A}}}\Psi(\delta_x)\text{tr}[\Omega\Phi(x)]=\bigoplus_a\Psi_a(\theta)$, where $\theta=\sum_{x\in B_{\mathcal{A}}}\delta_x \text{tr}[\Omega\Phi(x)]$ is known as the canonical regular element of $\mathcal{A}^*$.  Since $E$ is the transfer matrix of the MPDO, the fixed-point condition implies $E^2 = E$. Consequently, $\theta$ is an idempotent element in $\mathcal{A}^*$: $\Psi_a(\theta)^2=\Psi_a(\theta)$ for all $a\in\mathcal{B}$. Now collect all irreps $a$ of $\mathcal{A}^*$ such that $\Psi_a(\theta)\neq 0$, and name this set $\mathcal{I}$. The trivial representation $I$ must be in $\mathcal{I}$: $O_I^{(L)}=O_I^{(L)\dagger}O_I^{(L)}\ge 0$, hence $\text{tr}[O_I^{(L)}\Omega^{\otimes L}]=\text{tr}\Psi_I(\theta)>0$. We can define a modified tensor $M'=\sum_{x\in B_{\mathcal{A}}}[\Omega\Phi(x)]\otimes [\bigoplus_{a\in \mathcal{I}}\Psi_a(\delta_x)]$ so that $\rho^{(L-1)}(M,\mathcal{X}E)=\rho^{(L-1)}(M',\mathcal{X}')$ with $\mathcal{X}'$ the projection of $\mathcal{X}E$ onto irreps in $\mathcal{I}$, i.e., $\mathcal{X}'=\bigoplus_{a\in \mathcal{I}}\mathcal{X}_a \Psi_a(\theta)$. This is a renormalization fixed point generated by a simple tensor $M'$~\cite{Cirac2017}, meaning that its canonical form consists only of normal tensors that do not vanish after tracing over the physical indices. Graphically, these normal tensors are
\begin{equation}
\label{eqn:Ma-tensor}
M_a=\sum_{x\in B_{\mathcal{A}}} [\Omega\Phi(x)]\otimes \Psi_a(\delta_x)=
\begin{tikzpicture}[scale=0.6]
    \draw[thick,Red] (0,0)--(1,0);
    \filldraw[black, ultra thin] (0.5,0) circle[radius=0.15];
    \draw[thick] (0.5,-0.5)--(0.5,0.75);
    \filldraw[black, ultra thin] (0.4,0.4) rectangle (0.6,0.6);
    \node at (0.1,0.5) {\small $\Omega$};
    \node [anchor=center, font=\footnotesize, text=Red]  at (-0.3,0) {\small $a$};
\end{tikzpicture}\quad \text{for }a\in \mathcal{I},
\end{equation}
where we have used the notation $\sum_{x\in B_{\mathcal{A}}}\Phi(x)\otimes \Psi_a(\delta_x)=\begin{tikzpicture}[scale=0.6]
    \draw[thick,Red] (0,0)--(1,0);
    \filldraw[black, ultra thin] (0.5,0) circle[radius=0.15];
    \draw[thick] (0.5,-0.5)--(0.5,0.5);
    \node [anchor=center, font=\footnotesize, text=Red]  at (-0.3,0) {\small $a$};
\end{tikzpicture}$, and tracing over the physical indices yield $\Psi_a(\theta)|_{a\in\mathcal{I}}\neq 0$. According to Theorem 4.9 in \cite{Cirac2017}, the fixed-point condition for MPDO generated by simple tensors further implies that normal tensors $M_a |_{a\in \mathcal{I}}$ must be supported on mutually orthogonal physical subspaces. Therefore if there exists other normal tensor $M_a |_{a\in \mathcal{I}, a\neq I}$, its multiplication with $M_I$ in the physical space, i.e.,
\begin{equation}
\begin{tikzpicture}[scale=0.6]
    \draw[thick,Red] (0,1)--(1,1);
    \filldraw[black, ultra thin] (0.5,1) circle[radius=0.15];
    \draw[thick] (0.5,0.5)--(0.5,1.75);
    \filldraw[black, ultra thin] (0.4,1.4) rectangle (0.6,1.6);
    \node at (0.1,1.5) {\small $\Omega$};
    \node [anchor=center, font=\footnotesize, text=Red]  at (-0.3,1) {\small $a$};
    \draw[thick,Red] (0,0)--(1,0);
    \filldraw[black, ultra thin] (0.5,0) circle[radius=0.15];
    \draw[thick] (0.5,-0.5)--(0.5,0.75);
    \filldraw[black, ultra thin] (0.4,0.4) rectangle (0.6,0.6);
    \node at (0.1,0.5) {\small $\Omega$};
    \node [anchor=center, font=\footnotesize, text=Red]  at (-0.3,0) {\small $I$};
\end{tikzpicture}
=\sum_{x\in B_{\mathcal{A}}}[\Omega^2\Phi(x)]\otimes [\Psi_a\boxtimes\Psi_I](\delta_x)
\end{equation}
must vanish. However, this cannot be true since $\Psi_a\boxtimes\Psi_I\cong \Psi_a$ and the right hand side vanishes only when $M_a$ vanishes. Therefore, there is no additional normal tensor in the canonical form of $M'$ other than $M_I$. This implies that the reduced density matrix after tracing out one site, $\rho^{(L-1)}(M',\mathcal{X}')=\frac{1}{\text{tr}\Psi_I(\theta)}O_I^{(L-1)}\Omega^{\otimes (L-1)}$, is independent of the choice of boundary matrix $\mathcal{X}$. Hence, different choices of $\mathcal{X}=\Pi_m$ yield locally indistinguishable states with $\rho_0^{(L)}\sim O_I^{(L)}\Omega^{\otimes L}$. Furthermore, since $\rho^{(L-1)}(M',\mathcal{X}')$ is a normal MPDO, its transfer matrix $\Psi_I(\theta)$ must have a unique eigenvalue of largest magnitude and is therefore a rank-$1$ projector. Since we have shown that $\Psi_a(\theta) = 0$ for all $a \neq I$, the transfer matrix $E = \bigoplus_a \Psi_a(\theta)$ of the original MPDO $\rho^{(L)}(M,\mathcal{X})$ reduces to a rank-$1$ projector, implying that the correlation length is zero.

\subsection{$L$-independent normalization factor $\mathcal{N}_m$}
\label{subsec:normalization}
The above proof also implies the normalization factor $\mathcal{N}_m$ is $L$-independent. Let us write $\Pi_m=\sum_a \Pi_m^{(a)}P_a$.  We can take the trace of our constructed MPDO fixed points:
\begin{equation}
\begin{split}
\text{tr}[\rho_{\text{RFP},m}^{(L)}]&=\frac{1}{\mathcal{N}_m}\text{tr}[O^{(L)}(\Pi_m)\Omega^{\otimes L}]\\
&=\frac{\Pi_m^{(I)}}{\mathcal{N}_m}\text{tr}[O^{(L)}_I\Omega^{\otimes L}]=\frac{\Pi_m^{(I)}\text{tr}\Psi_I(\theta)}{\mathcal{N}_m}=1.
\end{split}
\end{equation}
Hence the normalization factor $\mathcal{N}_m=\Pi_m^{(I)}\text{tr}\Psi_I(\theta)$ is a $L$-independent factor.

\section{Example of $\text{Vec}_{G}^\omega$}
\label{sec:group_cohomology}
For $\text{Vec}_{G}^\omega$, the simple objects are labeled by group elements and we have the following $F$-symbols:
\begin{equation}
\begin{array}{c}
\begin{tikzpicture}[scale=.5, baseline={([yshift=0ex]current bounding box.center)}, thick]
\draw (0,0) -- (0,1);
\draw (0,1) -- (-2,3);
\draw (-1,2) -- (0,3);
\draw (0,1) -- (2,3);
\draw (0,-0.5) node{$g_1 g_2 g_3$};
\draw (-2,3.5) node{$g_1$};
\draw (0,3.5) node{$g_2$};
\draw (2,3.5) node{$g_3$};
\end{tikzpicture}
\end{array}
=\omega(g_1,g_2,g_3)
\begin{array}{c}
\begin{tikzpicture}[scale=.5, baseline={([yshift=0ex]current bounding box.center)}, thick]
\draw (0,0) -- (0,1);
\draw (0,1) -- (2,3);
\draw (1,2) -- (0,3);
\draw (0,1) -- (-2,3);
\draw (0,-0.5) node{\small $g_1 g_2 g_3$};
\draw (-2,3.5) node{\small $g_1$};
\draw (0,3.5) node{\small $g_2$};
\draw (2,3.5) node{\small $g_3$};
\end{tikzpicture}
\end{array}.
\end{equation}
From here we can construct the resulting $C^*$-weak Hopf algebra using the diagramatic approach in Sec.~\ref{subsec:reconstruction}. Similar reconstruction can be found in \cite{Cordova2024}. For convenience, we choose the 3-cocycle $\omega$ to be normalized, i.e., $\omega(g_1,g_2,g_3)=1$ if some of $\{g_1,g_2,g_3\}$ is the unit of the group. Let us construct the $C^*$-weak Hopf algebra $\mathcal{A}_{\text{Vec}_{G}^{\omega}}$ such that $\text{Rep}(\mathcal{A}^*_{\text{Vec}_{G}^{\omega}})=\text{Rep}(\mathcal{A}_{\text{Vec}_{G}^{\omega}})=\text{Vec}_G^{\omega}$: The algebra is spanned by basis $e_{g_1,g_2}^{(h)}$, where $g_1, g_2, h\in G$ and has a dimension of $|G|^3$ with $|G|$ being the order of the group. The multiplication is defined as
\begin{equation}
e_{g_1,g_2}^{(h_1)}\cdot e_{g_3,g_4}^{(h_2)}=\frac{\omega(h_1,h_2,g_3)}{\omega(h_1,h_2,g_4)}\delta_{g_1,h_2g_3}\delta_{g_2,h_2g_4}e_{g_3,g_4}^{(h_1h_2)},
\end{equation}
and the comultiplication is defined as
\begin{equation}
\Delta[e_{g_1 g_2}^{(h)}]=\sum_{g'} e^{(h)}_{g_1 g'}\otimes e^{(h)}_{g'g_2}.
\end{equation}
The unit, counit, antipode, and *-operation are as follows:
\begin{equation}
\begin{split}
&{\bf 1}=\sum_{g_1,g_2} e_{g_1,g_2}^{(I)},\quad \epsilon(e_{g_1,g_2}^{(h)})=\delta_{g_1,g_2},\\
&S(e_{g_1,g_2}^{(h)})=\frac{\omega(h^{-1},h,g_1)}{\omega(h^{-1},h,g_2)} e_{hg_2,hg_1}^{(h^{-1})},\\
& [e_{g_1,g_2}^{(h)}]^*=\frac{\omega(h^{-1},h, g_2)}{\omega(h^{-1},h,g_1)} e_{h g_1,h g_2}^{(h^{-1})}.
\end{split}
\end{equation}
The faithful *-representations for the algebra and dual algebra can be chosen as
\begin{equation}
\begin{split}
&\Phi[e_{g_1,g_2}^{(h)}]=\omega^{-1}(h, g_1,g_1^{-1}g_2)|hg_1,h g_2\rangle \langle g_1,g_2|, \\
&\Psi[\tilde{e}^{(h)}_{g_1,g_2}]=|h,g_1)(h, g_2|,
\end{split}
\end{equation}
where $\tilde{e}^{(h)}_{g_1,g_2}$ is the dual of $e^{(h)}_{g_1,g_2}$ and the matrix multiplication of $\Phi$ is consistent with the algebraic multiplication due to the cocycle condition. We have also used the Dirac notation to represent matrices on basis $|g_1,g_2\rangle$ and $|g_1,g_2)$ for the physical and virtual spaces, respectively. Both $\Phi$ and $\Psi$ furnish complete decompositions into the $|G|$ irreducible representations of $\mathcal{A}^*_{\text{Vec}_G^{\omega}}$ and $\mathcal{A}_{\text{Vec}_G^{\omega}}$, respectively, with each irrep appearing once. The representation $\Psi$ contains irreps labeled by $h$, which we denote as $\Psi_h$: $\Psi_h[\tilde{e}_{g_1g_2}^{(h')}]=\delta_{hh'}|h,g_1)(h, g_2|$. Correspondingly, we can obtain $O_h^{(L)}$, generated by the tensor $\sum_{h',g_1,g_2}\Phi[e_{g_1,g_2}^{(h')}]\otimes \Psi_h[\tilde{e}^{(h')}_{g_1,g_2}]$, satisfying the multiplication rule of group $G$:
\begin{equation}
O_{h_1}^{(L)}O_{h_2}^{(L)}=O_{h_1 h_2}^{(L)},
\end{equation}
and from the formalism developed in the main text, the MPDO fixed-point can be assembled as:
\begin{equation}
\rho_{\text{RFP},m}^{(L)}=\frac{1}{\mathcal{N}_m}O^{(L)}(\Pi_m)\Omega^{\otimes L},
\end{equation}
where $\Omega=\mathbb{1}_{|G|^2}/|G|$ since $\text{Rep}\mathcal{A}_{\text{Vec}_G^{\omega}}\cong\text{Vec}_G^{\omega}$ in our construction and every irrep has quantum dimension $1$.  Given any 1D irrep $m$ of the group $G$, the explicit form of $\Pi_m$ is
\begin{equation}
\Pi_m=\frac{1}{|G|}\sum_{h\in G}\chi_m(h^{-1})P_h,
\end{equation}
where $\chi_m$ is the character of the irrep $m$ of the group, and $P_h$ is the Hermitian projector from $\Psi$ onto irrep $\Psi_h$.
The resulting MPDO fixed points have a strong MPO symmetry $\{O_h^{(L)}\}_{h\in G}$ that are anomalous due to nontrivial 3-cocycle and are topologically nontrivial for similar reason as discussed in \cite{Lessa2025a}.

\section{Example of Fibonacci fusion category}
\label{sec:Fibonacci}
Here we give details about the $C^*$-weak Hopf algebra from Fibonacci fusion category $\mathcal{C}_{\text{Fib}}$. As discussed in the main text, the reconstructed $C^*$-weak Hopf algebra $\mathcal{A}_{\text{Fib}}$ is isomorphic to $\mathcal{M}_2(\mathbb{C})\otimes\mathcal{M}_3(\mathbb{C})$. One can choose the basis $e_{\alpha,ij}$ where for $\alpha=I$, $i,j\in\{1,2\}$ and for $\alpha=\tau$, $i,j\in\{1,2,3\}$. The multiplication follows the rules of matrix multiplication, 
\begin{equation}
e_{\alpha,ij}e_{\alpha',kl}=\delta_{\alpha \alpha'}\delta_{jk}e_{\alpha,il}.
\end{equation}
The comultiplication is defined as follows~\cite{Bohm1996}:
\begin{equation}
\begin{split}
\Delta(e_{I,11}) &:= e_{I,11} \otimes e_{I,11} + e_{\tau,11} \otimes e_{\tau,22},\\
\Delta(e_{I,12}) &:= e_{I,12} \otimes e_{I,12} 
  + \zeta^2\,e_{\tau,12} \otimes e_{\tau,21} 
  +\zeta  e_{\tau,13} \otimes e_{\tau,23},\\
\Delta(e_{I,22}) &:= e_{I,22} \otimes e_{I,22}
  + \zeta^4\,e_{\tau,22} \otimes e_{I,11}+\\
  &\zeta^3\,e_{\tau,23} \otimes e_{\tau,13}
  + \zeta^3\,e_{\tau,32} \otimes e_{\tau,31}
  + \zeta^2\,e_{\tau,33} \otimes e_{\tau,33},\\
\Delta(e_{\tau,11}) &:= e_{I,11} \otimes e_{\tau,11}
  + e_{\tau,11} \otimes e_{I,11}
  + e_{\tau,11} \otimes e_{\tau,33},\\
\Delta(e_{\tau,12}) &:= e_{I,12} \otimes e_{\tau,12}
  + e_{\tau,12} \otimes e_{I,21}
  + e_{\tau,13} \otimes e_{\tau,32},\\
\Delta(e_{\tau,13}) &:= e_{I,12} \otimes e_{\tau,13}
  + e_{\tau,13} \otimes e_{\tau,22}+\\
  & \zeta\,e_{\tau,12} \otimes e_{\tau,31}
  - \zeta^2\,e_{\tau,13} \otimes e_{\tau,33},\\
\Delta(e_{\tau,22}) &:= e_{I,22} \otimes e_{\tau,22}
  + e_{\tau,22} \otimes e_{I,11}
  + e_{\tau,33} \otimes e_{\tau,22},\\
\Delta(e_{\tau,23}) &:= e_{I,22} \otimes e_{\tau,23}
  + e_{\tau,23} \otimes e_{\tau,12}
  + \zeta e_{\tau,32} \otimes e_{\tau,21}-\\
  &\zeta^2\,e_{\tau,33} \otimes e_{\tau,23},\\
\Delta(e_{\tau,33}) &:= e_{I,22} \otimes e_{\tau,33}
  + e_{\tau,33} \otimes e_{I,22}
  + \zeta^2 e_{\tau,22} \otimes e_{\tau,11}-\\
  &\zeta^3\,e_{\tau,23} \otimes e_{\tau,13}
  - \zeta^3\,e_{\tau,32} \otimes e_{\tau,31}
  + \zeta^4\,e_{\tau,33} \otimes e_{\tau,33},
\end{split}
\end{equation}
with $\zeta=[(\sqrt{5}-1)/2]^{1/2}$. The comultiplication of remaining elements can be obtained from the *-operation which is antilinear and cohomomorphism:
\begin{equation}
e_{\alpha,ij}^*=e_{\alpha,ji}.
\end{equation}
The unit, counit, and antipode are defined as
\begin{equation}
\begin{split}
&{\bf 1}=\sum_{\alpha,i}e_{\alpha,ii},\\
&\epsilon(e_{I,ij})=1,\\
&\epsilon(e_{\tau,ij})=0,\\
&S(e_{I,ij})=e_{I,ji},\\
&S(e_{\tau,ij})=\zeta^{\xi(i)-\xi(j)} e_{\tau,\pi(j)\pi(i)},
\end{split}
\end{equation}
where $\pi,\xi$ are permutations of the set $\{1,2,3\}$ with $\pi$ swaps $1\leftrightarrow 2$ and $\xi$ swaps $2\leftrightarrow 3$. 

There exists a pairing
\begin{equation}
\langle, \rangle: \mathcal{A}_{\text{Fib}}\otimes\mathcal{A}_{\text{Fib}}\rightarrow \mathbb{C},
\end{equation}
which is a non-degenerate bilinear form that satisfies~\cite{Bohm1999}:
\begin{equation}
\begin{split}
&\langle x\cdot x', y\rangle=\langle x\otimes x',\Delta(y)\rangle,\quad\quad \langle {\bf 1}, y\rangle=\epsilon(y),\\
&\langle x, y\cdot y'\rangle=\langle \Delta(x), y\otimes y'\rangle,\quad\quad \langle x, {\bf 1}\rangle=\epsilon(x),\\
& \langle S(x), y\rangle=\langle x, S(y)\rangle,\quad \text{and}\quad \langle x^*, y\rangle=\overline{\langle x, [S(y)]^*\rangle},
\end{split}
\end{equation}
where the overline denotes the complex conjugation. The pairing induces an isomorphism $\mathcal{A}_{\text{Fib}}\cong \mathcal{A}_{\text{Fib}}^*$. In terms of basis, let us denote
\begin{equation}
\langle e_{\alpha',ij}, e_{\alpha,kl}\rangle=\tilde{R}_{\alpha,kl}^{\alpha',ij},
\end{equation}
and the isomorphism maps $e_{\alpha,ij}$ to $\sum_{\alpha',kl}\tilde{R}_{\alpha',kl}^{\alpha,ij}\tilde{e}_{\alpha',kl}\in \mathcal{A}_{\text{Fib}}^*$. We have a faithful *-representation of $\mathcal{A}$: $\Phi(e_{\alpha,ij})=|\alpha,ij\rangle\langle \alpha,ij|$. From the pairing-induced isomorphism, we can construct a faithful *-representation $\Psi$ for $\mathcal{A}_{\text{Fib}}^*$ satisfying
\begin{equation}
\Psi(\sum_{\alpha,kl}\tilde{R}_{a,kl}^{\alpha,ij}\tilde{e}_{a,kl})=|a,i)(a,j|,
\end{equation}
where we used notations introduced in the main text. Viewing $\tilde{R}_{a,kl}^{\alpha,ij}$ as a matrix of combined indices $\{\alpha,i,j\}$ and $\{a,k,l\}$, the $R$ matrix in the main text is the inverse of $\tilde{R}$ and
\begin{equation}
\Psi(\tilde{e}_{\alpha,ij})=\sum_{a,kl}R_{a,kl}^{\alpha,ij}|a,k)(a,l|.
\end{equation}
The construction for $\tilde{R}_{a,kl}^{\alpha,ij}$ is:
\begin{equation}
\begin{split}
\tilde{R}_{I,11}^{I,11}&=\tilde{R}_{I,12}^{\tau,11}=\tilde{R}_{\tau,11}^{I,12}=\tilde{R}_{I,21}^{\tau,22}=\tilde{R}_{\tau,22}^{I,21}\\
			      &=\tilde{R}_{\tau,21}^{\tau,21}=\tilde{R}_{\tau,23}^{\tau,31}=\tilde{R}_{\tau,31}^{\tau,23}=1,\\
\tilde{R}_{\tau,32}^{\tau,32}&=\tilde{R}_{\tau,13}^{\tau,13}=\zeta^{-1},\quad \tilde{R}_{\tau,12}^{\tau,12}=\zeta^{-2},\\
\tilde{R}_{\tau,33}^{I,22}&=\tilde{R}_{I,22}^{\tau,33}=-\tilde{R}_{\tau,33}^{\tau,33}=\zeta^2,\quad \tilde{R}_{I,22}^{I,22}=\zeta^4,\\
\end{split}
\end{equation}
and any other unspecified components are zero. The $R$ matrix as its inverse has its nonzero components:
\begin{equation}
\begin{split}
R_{I,11}^{I,11}&=R_{I,12}^{\tau,11}=R_{\tau,11}^{I,12}=R_{I,21}^{\tau,22}=R_{\tau,22}^{I,21}=R_{I,22}^{I,22}\\
&=R_{I,22}^{\tau,33}=R_{\tau,33}^{I,22}=R_{\tau,21}^{\tau,21}=R_{\tau,23}^{\tau,31}=R_{\tau,31}^{\tau,23}=1,\\
R_{\tau,13}^{\tau,13}&=R_{\tau,32}^{\tau,32}=\zeta,\quad R_{\tau,12}^{\tau,12}=-R_{\tau,33}^{\tau,33}=\zeta^2.
\end{split}
\end{equation}
From the constructions of $\Phi$ and $\Psi$, one can obtain the MPO tensor $\sum_{\alpha,ij}\Phi(e_{\alpha,ij})\otimes\Psi(\tilde{e}_{\alpha,ij})$. The result can be alternatively derived from the $F$-symbols of Fibonacci fusion category, see e.g.,~\cite{Molnar2022}. Our derivation naturally ensures $\Psi$ to be a faithful *-representation of $\mathcal{A}^*_{\text{Fib}}$.

In our construction, there are two irreps $I$ and $\tau$, whose quantum dimensions are $d_I=1$ and $d_{\tau}=(\sqrt{5}+1)/2$, and the projections to them is already trivially encoded in our construction. Let us now obtain the $O^{(L)}(\Pi_m)$ as MPO representations of the central idempotents corresponding to 1D irreps of Fibonacci fusion algebra. The nonzero components of fusion matrices are $N_{II}^{I}=N_{I\tau}^{\tau}=N_{\tau I}^{\tau}=N_{\tau \tau}^{I}=N_{\tau \tau}^{\tau}=1$. From the regular representation, we obtain two 1D irreps of the fusion algebra $\mathcal{R}$ and there are two corresponding central idempotents which should sum to the unit of the algebra. In MPO representions we have:
\begin{equation}
O^{(L)}(\Pi_0)+O^{(L)}(\Pi_1)=O^{(L)}_I.
\end{equation}
Since we know 
\begin{equation}
O^{(L)}(\Pi_0)=\frac{1}{\mathcal{D}^2}(d_I O_I^{(L)}+d_{\tau} O_{\tau}^{(L)})=\frac{2}{5+\sqrt{5}}O_I^{(L)}+\frac{1}{\sqrt{5}}O_{\tau}^{(L)},
\end{equation}
we must have
\begin{equation}
O^{(L)}(\Pi_1)=\frac{2}{5-\sqrt{5}}O_I^{(L)}-\frac{1}{\sqrt{5}}O_{\tau}^{(L)}.
\end{equation}
These equations give the result of Eq.~(20) in the main text. The corresponding symmetry eigenvalues are $\lambda_{0I}=1$, $\lambda_{0\tau}=\frac{\sqrt{5}+1}{2}$, and $\lambda_{1I}=1$, $\lambda_{1\tau}=\frac{-\sqrt{5}+1}{2}$.

\end{document}